# Co-evolution of primitive methane-cycling ecosystems and early Earth's atmosphere and climate

Boris Sauterey[1,2,3*], Benjamin Charnay[4], Antonin Affholder[1,2,3], Stéphane Mazevet[3,6], Régis Ferrière[1,2,5,6]

[1]Institut de Biologie de l'Ecole Normale Supérieure (IBENS), Université Paris Sciences et Lettres, CNRS, INSERM, 75005 Paris, France

[2]International Center for Interdisciplinary Global Environmental Studies (iGLOBES), CNRS, ENS-PSL University, University of Arizona, Tucson, AZ 85721, USA

[3]Institut de Mécanique Céleste et de Calcul des Ephémérides (IMCCE), Observatoire de Paris, université PSL, CNRS, Sorbonne Université, Univ. Lille, F-75014 Paris, France

[4]LESIA, Observatoire de Paris, Université PSL, CNRS, Sorbonne Université, Université de Paris, 5 place Jules Janssen, 92195 Meudon, France

[5]Department of Ecology & Evolutionary Biology, University of Arizona, Tucson, AZ 85721, USA

[§]These authors jointly supervised this work: Stéphane Mazevet, Régis Ferrière

*Correspondence to: boris.sauterey@ens.fr

**Abstract**

**The history of the Earth has been marked by major ecological transitions, driven by metabolic innovation, that radically reshaped the composition of the oceans and atmosphere. The nature and magnitude of the earliest transitions, hundreds of million years before photosynthesis evolved, remain poorly understood. Using a novel ecosystem-planetary model, we find that pre-photosynthetic methane-cycling microbial ecosystems are much less productive than previously thought. In spite of their low productivity, the evolution of methanogenic metabolisms strongly modifies the atmospheric composition, leading to a warmer but less resilient climate. As the abiotic carbon cycle responds, further metabolic evolution (anaerobic methanotrophy) may feed back to the atmosphere and destabilize the climate, triggering a transient global glaciation. Although early metabolic evolution may cause strong climatic instability, a low CO:CH4 atmospheric ratio emerges as a robust signature of simple methane-cycling ecosystems on a globally reduced planet such as the late Hadean/early Archean Earth.**

## Introduction

By 3.5 Ga, life had emerged on Earth[1-3]. Astrophysical and geophysical data concur in showing that the planet was habitable 400 My earlier at the very least, and possibly as early as ~4.5 Gya, depending on the occurrence, magnitude, and effect of large asteroid impacts during the Hadean[3]. Early on, Earth's carbon cycle likely established and maintained temperate climatic conditions[4,5] in spite of a Sun being 20-25% dimmer than it is today[6]. The earliest microbial ecosystems evolving under these conditions, hundreds of million years before the first anoxygenic phototrophs[7] became actors of the Archean climate[8], most likely involved chemolithotrophs (i.e., unicellular organisms that use redox potential as energy source for biomass production) producing methane as a metabolic waste. Phylogenetic analyses[2,7,9,10] combined with isotopic evidence[11] and the then-time predominance[12,13] of the electron donors $H_2$ and CO lend weight to a very early origin of $H_2$-based methanogens (MG), CO-based autotrophic acetogens (AG), and methanogenic acetotrophs (AT). As $CH_4$ built up in the atmosphere, the evolution of anaerobic methanotrophy (MT) may have been favored. Contrary to the modern biosphere, the biomass productivity of such ecosystems was likely low and energy limited, *i.e.*, limited by the availability of electron donors rather than nutrients such as nitrogen, phosphorus, or iron[13,14]. Whether and how the evolution of a primitive biosphere formed by these metabolisms influenced the planetary environment globally is unclear.

Previous studies[8,12,13] have addressed the productivity of primitive, chemolithotrophic ecosystems and their influence on the young Earth's equilibrium atmospheric conditions. Such studies relied on equilibrium analyses of the planetary ecosystem; they made strongly simplifying assumptions on the function of chemolithotrophic microbial metabolisms, and did not close the feedback loop linking biological activity, atmospheric composition, and climate. Although these studies showed that primitive biospheres may have had a significant impact on the planet's early atmosphere and climate, their ability to quantify this impact and estimate the underlying biomass productivity was limited. Furthermore, earlier theory based on equilibrium analyses could not address the coupled dynamics of metabolic evolution and planetary surface conditions, whereby evolutionary changes might trigger significant atmospheric and climatic events and lead to novel steady states. Thus, advancing existing models is needed to generate hypotheses on the history of atmospheric and climatic conditions that metabolic evolutionary innovation may have driven on the early Earth.

Here we ask, how constrained was the habitability of late Hadean/early Archean Earth to methane-cycling ecosystems? How productive were these ecosystems and did they have a significant impact on the atmospheric composition and climate? How did their impact change as different metabolisms evolved? To answer these questions, we lay out a new probabilistic modeling framework for an evolving microbial community coupled to early Earth surface geochemistry and climate (Fig. 1, see Methods and Supplementary Results and Discussion for further details). The mean surface temperature and the composition of the atmosphere and oceans are parameterized using a 3D climate model[4,15] and a 1D photochemical model[16], combined with a simple temperature-dependent carbon cycle model[5]. Advancing previous studies[8,12,13], this planetary model is coupled dynamically to a biological model of cell population dynamics and evolutionary adaptation, constructed by scaling the intracellular processes of energy acquisition (i.e., catabolism), cell maintenance, and biomass production (i.e., anabolism) up to ecosystem function[17,18]. The biological model is grounded in thermodynamics and based on observations of cell-size and temperature kinetic dependencies, widely and robustly shared among modern unicellular organisms[19-23].

Quantitative validation of similar models was obtained from laboratory experiments on anoxic ecosystems in bioreactors[18].

## Results

### *Ecosystems viability*

First we assess the viability of the methanogenic biospheres (MG, AG+AT, or MG+AG+AT) on an initially cool, lifeless Earth. The initial abiotic surface temperature $T_{Geo}$ is assumed to be 12°C, corresponding to $pCO_2 = 2\ 10^5$ ppm, negligible $pCH_4$, and volcanic outgassing of $H_2$, $\varphi_{volc}(H_2)$, ranging from $5\ 10^9$ to $2\ 10^{11}$ molecules $cm^{-2}\ s^{-1}$. By performing a Monte-Carlo exploration of the space of biological parameters, we generate posterior distributions of possible life-atmosphere-climate outcomes (Supplementary Figures 1 and 2), thus providing general insights that do not depend on specific parameterizations of chemolithotrophic metabolisms. We find that all three methanogenic ecosystems are viable (i.e., they can sustain a steady, positive biomass production) in more than 50 % of the simulations, regardless of the intensity of $H_2$ outgassing (further information on the region of viability in the space of biological parameters is provided in the Supplementary Results and Discussion). Among viable ecosystems, the posterior distributions of the planetary and ecological state variables are peaked and relatively narrow (Fig. 2). Hereafter we will focus on median values to describe model outputs.

### *Short-term effects of metabolic evolution*

We first consider the direct effects on the early Earth's atmosphere and climate, on a relatively short timescale of ~$10^6$ years, of the transition from the initially cool, lifeless state to a planet populated by one of three methanogenic biospheres, MG, AG+AT, or MG+AG+AT. On such a short timescale, the carbon cycle has a negligible influence on the atmospheric $CO_2$. $H_2$ is both a metabolic substrate of MG and involved in the photochemical production of CO, the metabolic substrate of AG (Fig. 1). As a consequence, stronger $H_2$ volcanic outgassing always enhances biomass production and $CH_4$ emission (Fig. 2). The highest $CH_4$ emission is achieved by MG ecosystems, with $pCH_4$ ranging from 80 to 4,000 ppm and equilibrium temperature, and $T_{BioGeo}$, raised by +7° to +17° (Fig. 2). The environmental impact of AG+AT ecosystems is similar in magnitude, with $pCH_4$ ranging from 50 to 1,000 ppm, and temperature increases of +6° to +15° (Fig. 2).

The planet's abiotic surface temperature, $T_{Geo}$, is likely to have a strong influence on methanogenic activity (see Methods). Temperature influences both cell kinetics (metabolisms are slower at lower temperature) and thermodynamics (strong negative effect of high temperatures on MG, less so on AG+AT). $T_{Geo}$ also correlates positively with $pCO_2$ and $pCO$, which are both substrates of methanogenesis. We evaluate the influence of $T_{Geo}$ by examining bio-geo environmental feedbacks (*i.e.*, how the change in metabolic activity due to variation in $T_{Geo}$ feeds back to climate) for $T_{Geo}$ ranging from -18 to 57 °C, which corresponds to $pCO_2$ ranging from $5\ 10^{-4}$ to 1 bar (Fig. 3). We ran simulations using a default biological parameterization for which $CH_4$ emissions are close to the median predictions described above. Note that we also considered $T_{Geo}$ as varying independently of $pCO_2$, due to *e.g.* variation in stellar radiation; corresponding results are shown in Supplementary Figure 3.

If the overall effect of $T_{Geo}$ on methanogenic activity is positive, the methanogenic ecosystem is expected to amplify temperature fluctuations driven by external events such as variation in $CO_2$ outgassing. In this case, increasing $T_{Geo}$ should enhance the biogenic emission of $CH_4$, further warming the planet through additional greenhouse effect; decreasing $T_{Geo}$ should have the opposite effect. In contrast, if the overall effect of $T_{Geo}$ on methanogenic activity is negative, methanogenic ecosystems will buffer temperature variation. For the MG ecosystem (Fig. 3), we find that there is a critical abiotic temperature $T_{Geo} \approx 5$ °C, almost insensitive to $H_2$ outgassing, at which the warming effect of the ecosystem is maximum, from +5° to +17° across the range of $H_2$ outgassing rates. The MG ecosystem buffers fluctuations of abiotic temperature above the critical value, whereas it amplifies abiotic temperature fluctuations below the critical value. In contrast, the AG+AT ecosystem always amplifies temperature variation at all abiotic temperatures, and its influence tends to dominate when MG and AG+AT metabolisms co-occur (Fig. 3). Such amplification or buffering of temperature variations can represent up to 33 % of the abiotic fluctuations (Supplementary Figure 4). Overall, the function and evolution of methanogenic ecosystems lead to a less resilient, more variable climate.

The formation of organic hazes when the $pCH_4$-to-$pCO_2$ ratio is greater than 0.2 (ref[24]) has been proposed as a general mechanism of climate regulation[25]. In the late Archean, $pCO_2$ was low[5], favoring organic haze formation that may have prevented hot runaway scenarios. In the Hadean/early Archean however, our model predicts organic hazes to form in a limited range of conditions, at very high $H_2$ volcanic outgassing rates, low $pCO_2$, and low abiotic temperature close to or below the freezing point (Fig. 3). Under these specific conditions, the formation of organic hazes may overwhelm the warming effect of methanogenic ecosystems and leave the planet in a globally glaciated state. Under most conditions however, it is the availability of electron donors ($H_2$, CO) to methanogenesis, and not organic hazes, that is expected to limit Archean climate warming by biological activity.

***Biomass production***

In spite of their strong impact on the planet's atmosphere and climate, primitive methanogenic ecosystems are characterized by extremely low biomass productivity, AG+AT being the most productive pathway by far (Fig. 2). As microbial chemolithotrophs consume atmospheric electron donors, they drive the system closer to its thermodynamic equilibrium, thus gradually decreasing the thermodynamic efficiency of the metabolic coupling between energy acquisition (catabolism) and biomass production (anabolism; see Methods equation (E2)). As a consequence, for the highest value of abiotic $H_2$ outgassing, our model predicts biomass production to range from $10^6$ to $10^9$ molecule C $cm^{-2}$ $s^{-1}$. This is 1 to 4 orders of magnitude below previous estimates[13] (based on models that assumed a fixed biomass yield per electron donor consumed) and 4 to 7 orders of magnitude below modern values[26].

Albeit extremely low, biomass production is very sensitive to the metabolic composition of the ecosystem and temperature, $T_{Geo}$. Supplementary Figure 5 shows how biomass production is influenced by these two factors. For the MG ecosystem, biomass production peaks for $T_{Geo}$ between -10 and 10 °C depending on the intensity of $H_2$ volcanic outgassing (slightly above $10^9$ molecules C $cm^{-2}$ $s^{-1}$ at high rates of $H_2$ volcanic outgassing), and strongly decreases for higher $T_{Geo}$. The maximum biomass production is of the same order in the AG+AT ecosystem and reached for similar conditions (intermediate $T_{Geo}$, high $H_2$ volcanic outgassing rate), but its dependence upon $T_{Geo}$ and the rate of $H_2$ volcanic outgassing is much weaker. In the MG+AG+AT ecosystem the two methanogenic pathways interact synergistically, leading to a nonlinear, multiplicative increase in biomass production at low and high temperature. The synergy involves the combination of biogeochemical recycling loops, both locally and globally. Locally, while MG consumes $CO_2$ to produce $CH_4$, AT decomposes $CH_3COOH$ and produces $CO_2$. The metabolic waste of AT is therefore the metabolic substrate of MG, and the combination of the two metabolic pathways pulls the system further away from its thermodynamic equilibrium, hence an increase in the efficiency of both pathways. Globally, as the MG metabolism releases additional $CH_4$ in the atmosphere, the production of CO through photochemistry is accelerated; CO being the metabolic substrate of AG, its metabolic efficiency is enhanced. The synergistic effect is greatest at low abiotic temperature. In those very specific conditions and for the highest values of $H_2$ outgassing, biomass production can reach about $10^{10}$ molecules C $cm^{-2}$ $s^{-1}$ -- about 1,000 times less than estimates of modern primary production[26].

### *Metabolic evolution and the carbon cycle*

Next, we investigate how the evolutionary process of metabolic diversification of the primitive biosphere shaped the planet atmospheric and climatic history. We consider alternate scenarios of biosphere evolutionary complexification (Fig. 4A) consistent with phylogenetic and geological inferences[2,7,9,10,11,27]. Our evolutionary sequences culminate with the evolution of anaerobic methanotrophy, which uses the oxidation of $CH_4$ as primary source of energy, based on the consumption of $H_2SO_4$, the main oxidative species of the globally reduced early Earth[28]. We therefore make the plausible assumption that a full methane-cycling biosphere may have evolved before the advent of photosynthesis.

This evolutionary process of diversification may have spanned several hundred million years (from the origin of life 3.9-4.5 Gya, to the origin of anoxygenic phototrophy, 3.5-3.7 Gya), thus unfolding on a timescale over which the geochemical cycles interacted dynamically and reciprocally with biological activity. In particular, it has been shown[5] that on the timescale of $10^7$ to $10^8$ years, the carbon cycle tends to mitigate temperature variations through a negative feedback on the atmospheric $CO_2$ concentration. We adapted the model from ref[5] to add a full, temperature-dependent carbon cycle to our planetary ecosystem model. The abiotic equilibrium $pCO_2$ is now determined by the balance between $CO_2$ outgassing and sequestration in the oceanic floor; $p$CH4, by the balance between serpentinization and photodissociation rates; and $pH_2$, by the balance between outgassing and photochemical reactions. Additionally the $H_2SO_4$ oceanic concentration is determined by the balance between rainout and hydrothermal remineralization rates (taken from ref[29]). By drawing values for the abiotic supplies in $CO_2$, $CH_4$, $H_2$, and $H_2SO_4$ from reasonable, log-uniform priors (Table 1), and starting from a lifeless primitive Earth, we evaluate the equilibrium state of the planetary system after each evolutionary metabolic transition in terms of atmospheric signatures (Fig. 4B, C and D) and mean surface temperature (Fig. 4E).

First, we find that a lifeless Earth is characterized by a high $CO:CH_4$ atmospheric ratio of $10^2$-$10^4$ to 1, which differs markedly from the ratio predicted with a functional biosphere, regardless of its metabolic composition. As the biosphere complexifies, biological activity increases the atmospheric concentration in $CH_4$, decreases the atmospheric CO, or both, causing the $CO:CH_4$ ratio to fall. By comparing median values between the abiotic state and the most complex biosphere (MG+AG+AT+MT), the $CO:CH_4$ ratio is predicted to be reduced by a factor of ~5,000. The earliest evolutionary events, whereby the MG, AG, or AG+AT ecosystem emerges, all cause atmospheric shifts that can be distinguished in the $p$CO-$p$H$_2$-$p$CH$_4$ space (Fig. 4B and C). The atmospheric shifts caused by subsequent evolutionary complexification (leading to MG+AG, MG+AG+AT, or MG+AG+AT+MT ecosystems) are less pronounced and the corresponding atmospheric signatures are less distinctive among themselves. The evolution of anaerobic methanotrophy (MT) has for instance no effect on the equilibrium atmosphere of the planet because the influx of $H_2SO_4$ is sufficient for methanotrophs to survive and co-occur with methanogens, but not for them to consume a significant portion of the $CH_4$ produced by methanogens.

Finally, although methanogenesis can have major effects on climate on relatively short time scales (Figs. 2 and 3), the carbon cycle buffers these effects at equilibrium on longer time scales: the average temperature difference between the planet with and without a methanogenic biosphere on its surface is only 4°C (Fig. 4D). Biological effects on temperature may be further attenuated by the formation of cooling organic hazes favored by the enhanced sequestration of $CO_2$ in response to methanogenesis[25]. However, the necessary conditions for organic hazes to form in this case are met in only 0.03% of the simulations including a methanogenic biosphere.

***Transient climate destabilization by methanotrophy***

How the planetary atmosphere-climate system responds to metabolic innovation may depend on the pace of evolution itself. The evolution of methanotrophy is a case in point. Slow evolution may delay methanotrophy after the response of the carbon cycle to methanogenesis. In this case, the oceanic stocks of $H_2SO_4$ are sufficient for methanotrophs to rapidly consume almost all of the atmospheric $CH_4$ (Supplementary Figure 6) and the planet is temporarily characterized by an atmospheric deficit in both $CO_2$ and $CH_4$. As a result, temperature plummets below the initial abiotic temperature. In contrast, if evolution is fast enough and methanotrophs evolve before equilibration of the methanogenic biosphere with the planetary atmosphere and climate, atmospheric $CH_4$ may be consumed during or after the warming period (Fig. 3) but before the deficit in $pCO_2$ builds up; temperature then returns close to its initial value (Supplementary Figure 6).

Figure 5A illustrates the atmospheric and climatic consequences of slow evolution, when the wait time for methanotrophy is of the order of the carbon cycle timescale. In the two examples shown, the evolution of methanotrophs causes the sharp climatic response described above, and temperature falls respectively by 8 and 10 °C below $T_{Geo}$ = 2 °C, driving the planet into a global glaciation. Among 2,000 simulated planetary conditions characterized by randomly drawn abiotic characteristics (i.e., the abiotic influxes of $CO_2$, $H_2$, $CH_4$ and $H_2SO_4$, thereby setting the abiotic $pCO_2$, $pH_2$, $pCH_4$, initial oceanic concentration of $H_2SO_4$ and surface temperature $T_{Geo}$), and evolution time of methanotrophy (from 0 to 125 million years after methanogenesis), 50 % experience a temperature drop larger than 8.3 °C below $T_{Geo}$ (Fig. 5B), and 40% actually end up in glaciation (Fig. 5B and Supplementary Figure 7). As expected, delayed evolution of methanotrophy makes extreme cooling more likely (Fig. 5B). Because they lead to an enhanced abiotic response of the carbon cycle, conditions for which methanogenic ecosystems have the greatest warming effect (low abiotic $T_{Geo}$, high $H_2$ outgassing rate) are also the conditions under which the evolution of methanotrophy has the most dramatic effect on climate and habitability (Fig. 5A and C, Supplementary Figure 8). Noticingly, this transient occurs irrespective of the metabolic composition of the methanogenic biosphere prior to the evolution of methanotrophy, as illustrated in Supplementary Figure 7.

For a methane-cycling ecosystem that triggers and eventually survives global glaciation, we expect the resulting equilibrium (co-occuring methanogens and methanotrophs with low $pCH_4$) to be maintained only transitorily. We assume that during the early Archean, prior to the evolution of methanotrophy, the oceanic stock of $H_2SO_4$ builds up as volcanoes emit $SO_2$ photochemically transformed into $H_2SO_4$, which then deposits on the ocean surface. With a deposition rate corresponding to our lowest value of $10^7$ molecules cm$^{-2}$ s$^{-1}$ (ref[30]), and a hydrothermal removal rate of $[H_2SO_4]_{oc.} \times 7.2\ 10^{12}$ L y$^{-1}$ (ref[29]), we obtain an abiotic oceanic concentration of 0.4 mM. Such a stock is sufficient for methanotrophs to consume most of the atmospheric $CH_4$, leading to the global cooling described above. Since the rate of $CH_4$ reduction by methanotrophs is faster than the deposition rate, the $H_2SO_4$ stock will ultimately be depleted by methanotrophs, thereby driving a new abrupt environmental shift towards the equilibrium described in Fig. 4. At this new, stable equilibrium of coexisting methanogens and methanotrophs, the atmospheric $pCH_4$ is high and and the $H_2SO_4$ oceanic concentration is very low, below the micromolar, which aligns with the sulfur isotopic fractionation records[30-32]. This result suggests that anaerobic methanotrophy may have been the main sink of $H_2SO_4$ prior to the Neoarchean (2.5 to 2.7 Gya).

**Discussion**

Our initial questions were: How constrained was the habitability of late Hadean/early Archean Earth to methane-cycling ecosystems? How productive were these ecosystems and did they have a significant impact on the atmospheric composition and climate? How did their impact change as different metabolisms evolved? To address these questions, we build on previous theory[8,12,13] by setting up a probabilistic modeling framework in which the evolution of a microbial biosphere is coupled to the dynamics of early Earth surface geochemistry and climate. We focus on four simple microbial metabolisms involved in methane cycling and likely to be some of the earliest players in Earth's ecology: hydrogenotrophic methanogenesis (MG), acetogenesis (AG), methanogenic acetotrophy (AT), and anaerobic oxidation of methane (MT). The biosphere evolves when a metabolism that was not present appears and adapts. By closing the global feedback loop between biological and planetary surface processes, the model predicts both ecosystem and atmosphere-climate states under the assumption that they reciprocally influence one another.

Our results confirm the contention that the late Hadean/early Archean planet was most likely habitable to methane-cycling chemolithotrophic biospheres and that under the assumption of high enough $H_2$ supply, these biospheres were key factors of the climatic and atmospheric evolution of the planet[8,12,13,33,34]. On short time scales ($10^5$-$10^6$ years) the evolution of methanogenic biospheres may have considerably warmed the climate and influenced its resilience, in spite of a very low ecosystem productivity. On longer timescales, commensurate with the abiotic response of the carbon cycle to temperature variation, all ecosystems converge to new stable equilibria. Under these long-term equilibrium conditions, the mean surface temperature does not differ much from the lifeless state, and the carbon cycle is the predominant mechanism of climate regulation. However, all ecosystem equilibria share a robust atmospheric signature (low $CO:CH_4$ ratio), distinctive from the lifeless state. In addition to influencing planetary characteristics at equilibrium, metabolic evolution generates atmospheric and climatic transients. The pace of evolution thus has a strong influence on the atmosphere and climate history. In particular, fast evolution of methanotrophs has limited or no effect on climate, whereas their delayed evolution may cause strong transients leading to global glaciation.

Although the influence of chemotrophic methane-cycling ecosystems on climate has been discussed in previous work[8,12,13,33,34], our model is the first to couple models of the Archean atmosphere, climate, and carbon cycle to an explicit eco-evolutionary model of cell population dynamics in order to quantify this effect. Our model differs from previous work primarily by addressing how climate change driven by ecological function feeds back to the biological activity of microbial populations, through the thermal dependence of the thermodynamics and kinetics of cell metabolism, and triggers an abiotic response of the carbon cycle. Closing the global feedback loop between ecological and planetary processes allows us to predict the ecosystem and climate states under the assumption that they reciprocally influence one another on multiple timescales.

A general result is that MG and AG+AT ecosystems are characterized by extremely low biomass production relative to their planetary impact on the atmosphere and climate. We predict biomass production to be 1 to 4 orders of magnitude smaller than previous estimates[13] and 3 to 7 orders of magnitude below modern values[26]. Maximum global biomass production (reached for a very specific combination of biotic and abiotic conditions) is $10^{10}$ molecules C $cm^{-2}$ $s^{-1}$, or 3 $10^{7}$ ton C $year^{-1}$. This is approximately 1,000 times less than estimates of modern primary production. Both in terms of stock and fluxes, biomass has therefore very little effect on the biogeological coupling, which is a major difference with modern ecosystems.

A key assumption of our model is that nutrients N and P are not limiting; this assumption is backed up by the prediction of very low biomass production. As previously argued[13,14], primitive ecosystems with such low productivity should have been limited by the availability in electron donors rather than by N and P nutrients. More specifically, ref[14] evaluated the most likely limitation to biomass production prior to the evolution of oxygenic photosynthesis. By assuming a fixed yield for biomass production per molecules of electron donor consumed (as in ref[13]), they found that the N and P supplies were most likely sufficient during the Archean for electron donors to be limiting. This is even more likely given our finding of a biomass productivity far lower than the previous estimates[13,14].

Our prediction of very low biomass productivity may help better constrain the timeline of the evolution of metabolic innovation on Earth. Even in our most productive scenario and under the (extreme) assumption that 100% of the dead organic matter is buried before remineralization, the estimated biomass production is still 4 to 5 times lower than the level consistent with the C isotopic fractionation estimated from rocks as old as 3.5 Gy[1,35]. This suggests that more productive, likely photosynthetic life forms must have evolved more than 3.5 Gya ago.

By performing equilibrium analyses of the planetary system on longer time-scales ($10^7$-$10^8$ years), on which the carbon cycle responds to ecosystem function and sequential metabolic diversification of the biosphere, we find that the climate regulation of the planet by the abiotic carbon cycle largely buffers the influence of early methanogenic activity on climate. However, depending on the pace of evolution, metabolic transitions such as the evolution of methanotrophy can interact with the abiotic carbon cycle and trigger strong transitory climatic events such as global glaciation over $10^7$ to $10^8$ years.

Even though the timing of the origin of anaerobic oxidation of methane (AOM), by reduction of nitrate, nitrite, or sulfate, is not well resolved, the fact that AOM proceeds enzymatically as a reversal of methanogenesis has been used to suggest that AOM may have evolved relatively soon after methanogens, *i.e.* before phototrophy[36,37]. Additionally, we find that the evolution of AOM prior to photosynthesis appears to be compatible with the sulfur isotopic fractionation record[30-32]. Our results thus highlight biological activity and the evolution of pre-photosynthetic methane-cycling ecosystems as a potential destabilizing factor of the early Earth climate system, alternatively or additionally to abiotic causes such as large impactors[3,4,38].

The mechanism by which the evolution of early methane-cycling ecosystems may have exposed the Earth to high risks of global glaciation – fast removal of atmospheric methane, slow response of the carbon cycle – is general. It is therefore tantalizing to speculate about its potential involvement in the major climatic events that paralleled the evolution of oxygenic photosynthesis. By poisoning methanogens, removing the resource limitation (oxidizing species: $O_2$, $H_2SO_4$, $NO_3$, $NO_2$) of methanotrophs, or by simply driving the abiotic oxidation of atmospheric $CH_4$ by outgassed $O_2$, oxygenic photosynthesis may have driven a dramatic decline in atmospheric $CH_4$ which, coupled to a delayed response of the carbon cycle, could have caused global cooling, triggering the Proterozoic glaciation ~2.3 Gya[39-41].

The substantial shifts in atmospheric composition driven by evolutionary metabolic innovation (Fig. 4) highlight the fact that simple methane-cycling ecosystems can have radically different atmospheric signatures. Thus, the atmosphere composition is shaped very early on by the course of biological evolution. We observe that the atmospheric signature of a specific metabolism is highly dependent on the ecological context set by the whole metabolic composition of the biosphere. As the biosphere evolves and complexifies, the atmospheric signature of each metabolic type becomes less identifiable, yet the state of the planet remains distinctive from the lifeless scenario. This backs up and quantifies the hypothesis that on a globally reduced planet such as the early Archean Earth, a low $CO:CH_4$ atmospheric ratio is a highly discriminant indicator of inhabitation by a primitive $CH_4$-based biosphere[42].

In conclusion, our results suggest that life has had a dramatic impact on the anoxic Earth's surface environment, long before phototrophy evolved. They highlight the importance of the evolutionary process and its timeline in shaping up the planet's atmosphere and climate. In spite of extremely low productivity, metabolic evolutionary innovation in primitive methane-based biospheres is predicted to cause distinctive shifts in atmospheric composition, such as a decreasing $CO:CH_4$ ratio as greater metabolic complexity evolves. The warming effect of methanogens and cooling effect of methanotrophs can be strong but they are only transient, on timescales that depend on the pace of evolution. We anticipate that the continued development of models that couple planetary processes with ecological and evolutionary dynamics of microbial biospheres will further advance our understanding of major events in the co-evolutionary history of life and Earth, and help identify detectable biosignatures for the search of life on Earth-like exoplanets.

# METHODS

## Biological model

In the following section, we present the biological model. Parameters' definition, unit and default value are given in Supplementary Tables 1, 2, and 3.

The biological model describes the dynamics of one or several biological populations of chemotrophic organisms, driven by the growth, birth (division) and death of individual cells. The individual cell life cycle is controlled by catabolic and anabolic reactions occurring within cells[17,18,43]. Catabolism produces energy used by anabolism for biomass production, which determines cell growth and division; a fraction of energy produced by catabolism is used for cell maintenance. Energy thus flows from catabolism to maintenance and anabolism, and these processes can be described as follows:

$$\begin{gathered} Catabolism: \sum_{i=1}^{n} \gamma_{S_i}^{cat} S_i \rightarrow \sum_{i=1}^{m} \gamma_{P_i}^{cat} P_i + \Delta G_{cat} \\ Maintenance: E_m \\ Anabolism: \sum_{i=1}^{n'} \gamma_{S_i}^{cat} S_i + \Delta G_{ana} \rightarrow \sum_{i=1}^{m'} \gamma_{P_i}^{cat} P_i + B \end{gathered} \tag{E1}$$

where $S_i$ and $P_i$ are the substrates and products of the metabolic reactions, and the $\gamma$'s are the corresponding stoichiometric coefficients. We follow ref[17] and assume that the average organic compound $CH_{1.8}O_{0.5}N_{0.2}$ is a plausible approximation for the living biomass $B$. The energy $E_m$ (in kJ d$^{-1}$) measures the cost of maintenance for a single cell per unit of time. The $\Delta G$ terms measure the oxidative power released or consumed by the catabolic and anabolic reactions; they are given by the Nernst relationship:

$$\Delta G(T) = \Delta G_0(T) + RT\, log(\prod_{i=1}^{n} S_i^{\gamma_{S_i}} \prod_{i=1}^{m} P_i^{\gamma_{P_i}}) \tag{E2}$$

where $R$ is the ideal gas constant, $T$ is the temperature (in K) and $\Delta G_0(T)$ (in kJ) is the Gibbs energy of the reaction. $\Delta G_0(T)$ is obtained from the Gibbs-Helmholtz relationship:

$$\Delta G(T) = \Delta G_0(T_S)\frac{T}{T_S} + \Delta H_0(T_S)\frac{T_S - T}{T_S} \tag{E3}$$

where $T_S$ is the standard temperature of 298.15 K. Note that equations (E2) and (E3) describes how the thermodynamics of any given metabolism (*i.e.* the combination of catabolism and anabolism) vary with temperature.

Michaelis-Menten kinetics apply to catabolism:

$$q_{cat} = q_{max} \frac{min(S_i^{cat}/\gamma_{S_i}^{cat})}{min(S_i^{cat}/\gamma_{S_i}^{cat}) + K_S} \tag{E4}$$

where $K_S$ is the half-saturation constant, $q_{max}$ the maximum metabolic rate of the cell ($d^{-1}$), and $min(S_i^{cat}/\gamma_{S_i}^{cat})$ measures the concentration of the limiting substrate (taking stoichiometry into account). The energy produced is first directed toward maintenance. The cell energetic requirement for maintenance is:

$$q_m = \frac{-E_m}{\Delta G_{cat}} \tag{E5}$$

The cell therefore meets its energy requirement only if $q_{cat} > q_m$. If the cell does not meet this requirement, the basal mortality rate of the cell, $m$, (in $d^{-1}$), is augmented by a decay-related mortality term equal to $k_d\ (q_m - q_{cat})$. Hence the actual mortality rate:

$$\begin{array}{c} If\ q_{cat} > q_m\colon\ d = m \\ If\ q_{cat} < q_m\colon\ d = m + k_d\ (q_m - q_{cat}) \end{array} \tag{E6}$$

When the energy requirements for maintenance are not met ($q_{cat} < q_m$) no energy is allocated to biomass production. Conversely, when those requirements are met ($q_{cat} > q_m$), then the energy remaining after allocation to cell maintenance can be directed to anabolism. A constant quantity of energy$\Delta G_{diss}$ (in kJ) is then lost through dissipation. Following ref[17] we consider the following empirical relationship:

$$\Delta G_{diss} = 200 + 18\ (6 - NoC)^{1.8} + e^{(-0.2-\alpha)^{2.16}\ (3.6\ +\ 0.4\ NoC)} \tag{E7}$$

where $NoC$ is the number of carbons in the carbon source used by the anabolic reaction, and $\alpha$ is its degree of oxidation. The efficiency of metabolic coupling (*i.e.*, the number of occurrences of the catabolic reaction to fuel one occurrence of the anabolic reaction once the maintenance cost has been met) is then measured by

$$\lambda = \frac{-\Delta G_{cat}}{\Delta G_{ana} + \Delta G_{diss}} \tag{E8}$$

The Michaelis-Menten kinetics of anabolism is given by

$$If\ q_{cat} > q_m: q_{ana} = \lambda\ (q_{cat} - q_m) \frac{min(S_i^{ana}/\gamma_{S_i}^{ana})}{min(S_i^{ana}/\gamma_{S_i}^{ana})\ +\ K_S}$$
$$If\ q_{cat} < q_m: q_{ana} = 0 \qquad \text{(E9)}$$

where the half-saturation constant $K_S$ independent of the substrate, as was assumed in ref[43]. As biomass accumulates in the cell at rate $q_{ana}$, cell growth may lead to cell division. The cell division rate, *r*, is given by:

$$If\ B < 2B_{Struct}: \quad r = 0$$
$$If\ B > 2B_{Struct}: \quad r = r_{max} \frac{1}{1\ +\ e^{-\ \theta\ log10((B\ -\ 2B_{Struct})/B_{Struct})}} \qquad \text{(E10)}$$

This means that a cell cannot divide if its internal biomass is not at least twice its cell structural biomass so that the two daughter cells meet this structural requirement. Above this threshold value of $2B_{Struct}$, $r$ is of sigmoidal form so that the division rate first increases exponentially with the intracellular biomass content then saturates to $r_{max}$ when biomass is largely available.

Using the rates of catabolic and anabolic cell activity and resulting cell division and mortality rates, we derive the following system of ordinary differential equations driving the cell population dynamics (dynamics of the number of cells and average cell biomass) and the feedback of the population on its environment (chemical composition of the ocean surface):

$$\frac{dN_i}{dt} = (r_i - d_i)\ N_i$$
$$\frac{dB_i}{dt} = q_{ana_i} - r_i\ B \qquad \text{(E11)}$$
$$\frac{dX_j}{dt} = F(X_j) + \sum_{i=1}^{MT} (q_{cat_i}\ \gamma_{i,X_j}^{cat} + q_{ana_i}\ \gamma_{i,X_j}^{ana})\ N_i$$

where $MT$ denotes the ensemble of metabolic types considered, $N_i$ is the number of individual cells in a population of a given metabolic type, $B_i$ is the average cellular biomass of that type, and $X_1, \ldots, X_S$ are the concentrations of all relevant chemical species in the environment. The term $\sum_{i=1}^{MT}(q_{cat_i}\ \gamma_{i,X_j}^{cat} + q_{ana_i}\ \gamma_{i,X_j}^{ana})\ N_i$ describes how concentrations vary according to the biological activity of each biological population $i$ in the microbial community. The $F(X_j)$ terms describe the environmental forcing resulting from ocean circulation and atmosphere-ocean exchanges as simulated a stagnant boundary layer model as in ref[13]. The system of equations (E11) is solved numerically using a forward Euler method.

The flow of energy through the cell is driven by the maximum metabolic rate, $q_{max}$ (in $d^{-1}$), and the rate of energy consumption for maintenance, $E_m$ (in kJ $d^{-1}$). They are both expressed as functions of temperature and cell size of the form $e^{a\,+\,bT}V^c$. Default values for parameters $a$, $b$ and $c$ entering $q_{max}$ and $E_m$ are given in Supplementary Table 2.

The structural cell biomass $B_{Struct}$ increases with cell size according to $B_{Struct} = aV^b$. Both metabolic rate and maintenance cost increase with cell size, but not as fast as structural biomass. Consequently, the biomass specific rates of metabolism and energy consumption for maintenance decrease with cell size. Thus, small organisms are better at acquiring energy, but large organisms are more cost efficient due to lower maintenance requirements. A trade-off mediated by cell size thus exists between metabolic and maintenance rates, hence an optimal (intermediate) cell size, which is consistent with previous work conducted on unicellular marine organisms[22]. Both metabolic and maintenance rates increase with temperature, but the metabolic rate increases slightly faster[19,20], shifting the trade-off toward larger sizes. As a consequence, the optimal cell size increases with temperature.

To compute the evolutionarily optimal cell size for a given metabolic type at a given temperature, we run simulations across a range of cell size and measure the level of resource use at equilibrium given by $Q^* = \prod_{i=1}^{n} S_i^{\gamma_{S_i}^{cat}} \prod_{i=1}^{m} P_i^{\gamma_{P_i}^{cat}}$, the product of the metabolic substrates and wastes concentrations at equilibrium weighted by their stoichiometric coefficient. According to the classical principle of "pessimization" (maximization of resource use)[44-46], populations that are better at exploiting their environment have lower $Q^*$ and evolution by natural selection will favor the cell size that minimizes $Q^*$, thus leading to the evolutionarily optimal cell size. Supplementary Figure 9 shows $Q^*$ as a function of cell size and temperature for the $H_2$-based methanogenesis.

The optimal cell size, $s_c^*$, follows a positive relationship with temperature given by $s_c^* = 10^{a\,+\,bT}$. We consider that over geological time scales, adaptive evolution acting on genetic variation among cells is fast enough so that $s_c$ is equal to $s_c^*$. As temperature changes, evolutionary adaptation by natural selection tracks the temperature-dependent optimal cell size.

In contemporary Earth oceans, most of the dead biomass is recycled by fermentors (>99%), and the rest is buried into the ocean floor. Although the most general version of our model includes populations of acetogenic biomass fermentors and acetotrophs, their inclusion considerably increases simulation time (results not shown). All the results presented in the main text have been obtained without fermentors, assuming that the dead biomass accumulates in the ocean's interior. It appears that biomass productivity is so low that the effect of biomass recycling (or lack thereof) on the atmospheric composition is negligible. We verified this by testing the effect of a recycling pathway in each ecosystem (MG, AG+AT, MG+AG+AT), for an intermediate value of $H_2$ volcanic outgassing. Although biomass production is sufficient to sustain a population of fermentors, we found no significant effect of their biological activity on the atmospheric composition. Additionally, when we can compare the global atmospheric redox budget of the planet with and without biomass production (see Supplementary Results and Discussion, Supplementary Figure 12), we find no significant differences. This further demonstrates that biomass production and the accumulation of dead biomass in the absence of remineralization represents a sink of C and H that is marginal compared to the other fluxes in the model. Note that the situation would be very different after the evolution of more productive, photosynthetic primary producers.

Because we do not model the fate of dead biomass explicitly, we evaluate the consistency of our predictions of biomass production with the geological record (carbon isotopic fractionation data) by taking biomass production as the upper bound for burial (in which case 100% of the produced biomass is ultimately buried) and using the modern value of burial (taken to be 0.2% as in e.g. ref[13]) as lower bound.

**Planetary model**

Our computation of climate state and mean surface temperature is based on 3D simulations with the Generic LMD GCM[4,47,48]. The model includes a 2-layer dynamic ocean computing heat transport and sea ice formation[49]. The radiative transfer is based on the correlated-*k* methods with *k*-coefficients calculated using the HITRAN 2008 molecular database. We used the simulations described in ref[4] for the early Earth at 3.8 Ga, assuming no land, 1 bar of $N_2$ and a 14 hour rotation period. The simulation grid covers a range of $pCO_2$ from 0.01 to 1 bar and $pCH_4$ from 0 to 10 mbar. From this grid, we derived the following simple parametrization of the mean surface temperature as a function of $pCO_2$ and $pCH_4$ (expressed in bar):

$$T\ (°C) = -19.26 + 77.67\sqrt{pCO_2} + 5\,log10(1 + \frac{pCH_4\ 10^6}{3}) \qquad \text{(E12)}$$

In the coupled biological-planetary model, we assume that the climate is always at equilibrium, meaning that the timescale of climate convergence is shorter than biological and geochemical timescales.

Photochemistry is computed with the 1D version of the Generic LMD GCM, which now includes a photochemistry core from ref[50]. The chemical network includes 30 species (mostly hydrocarbons) and 114 reactions. We use the reactions from ref[50] for $H_2$-$H_2O$-$CO_2$ and from ref[51] for hydrocarbons. The photochemistry model also includes a pathway for the formation of hydrocarbon aerosol ($C_2H + C_2H_2 \rightarrow HCAER + H$)[51,52].

We use the eddy diffusion vertical profile from ref[53] and the solar UV spectrum at 3.8 Ga from ref[54]. Boundary conditions are set by the mixing ratios of $CO_2$, $CH_4$, and $H_2O$ at the first atmospheric layer. The atmospheric CO is either consumed biotically by acetogens when they are present in the biosphere, or abiotically by the formation of formate in the ocean (resulting in a deposition velocity of $\sim 10^8$ cm.s$^{-1}$) as in ref[13,55]. This eliminates the possibility of CO-runaway, which otherwise occurs in our simulations at high level of $H_2$ or $CO_2$. In addition, we assume that $H_2$ undergoes diffusion-limited atmospheric escape (see for instance ref[13]).

We performed simulations across a range of $p$CO$_2$ ($10^4$ – $10^5$ ppm), $p$CH$_4$ (1 – $10^4$ ppm), $H_2$ (100 – $10^4$ ppm). For our range of atmospheric composition, photochemistry is dominated by the photolysis of $CO_2$ and $CH_4$, whose net reactions are:

$$CO_2 + H_2 = CO + H_2O \quad \text{(R1)}$$

$$CH_4 + CO_2 = 2CO + 2H_2 \quad \text{(R2)}$$

From the simulation outputs we derived simple parametrizations for the production rates and loss rates of $CH_4$, CO, $H_2$ (see Supplementary Figure 10). The reaction rates of (R1) and (R2) can be parameterized respectively as (in molecules cm$^{-2}$ s$^{-1}$):

$$F_1 = 1.8\ 10^{10} \cdot \left(\frac{pCO_2}{10^{-1}}\right)^{0.5} \cdot \left(\frac{pH_2}{10^{-4}}\right)^{0.4} \quad \text{(E13)}$$

$$F_2 = 2\ 10^{10} \cdot \left(\frac{pCO_2}{10^{-1}}\right)^{-0.2} \cdot \left(\frac{pH_2}{10^{-4}}\right)^{-0.2} \cdot \left(\frac{pCH_4}{10^{-4}}\right)^{u} \quad \text{(E14)}$$

where $u$ (between 0.5 and 1) is given by

$$u = 0.6\ +\ 0.1\ log(\frac{pH_2}{10^{-4}}) - \frac{log(pCH_4) + 2}{15} + 0.08\ log(\frac{pCO_2}{10^{-1}}) \quad \text{(E15)}$$

To simulate the evolution of $pCO_2$ with time and feedbacks between the microbial community, climate, and the carbon cycle, we use a simple carbon cycle model based on ref[5]. The model computes the evolution of atmospheric $CO_2$, dissolved inorganic carbon ($CO_2$, $HCO_3^-$ and $CO_3^{2-}$) and pH in the ocean and in seafloor pores. The model takes into account outgassing (from volcanoes and mid-oceanic ridges), continental silicate weathering, dissolution of basalts in the seafloor, and oceanic chemistry, as sources and sinks of $CO_2$. For all parameters in the carbon cycle model, we use the mean values given in ref[5]. For the temperature dependences of silicate weathering and oceanic chemistry, we use the mean surface temperature from our climate model. Even though this model only computes global mean quantities and does not take into account the organic matter in carbon sources and sinks of carbon, it is computationally very fast and well suited for studying the major feedbacks between biological populations and the atmosphere and climate.

**Coupled biological-planetary model**

The model resolves dynamics that take place on extremely different timescales -- geological processes are extremely slow ($10^3$ to $10^6$ years) compared to biological dynamics (days to years). We use a timescale separation approach and assume that the planetary environment is fixed on the biological time scale, and the biological system (population and local environment) is at ecological and evolutionary equilibrium on the geological time scale. This allows us to resolve the biological and geological dynamics separately and couple them at discrete points in time.

The biological dynamics are first resolved by the biological model for a fixed environmental forcing (equations (E11)), with microbial cells at their evolutionarily optimal size. The biological model is integrated over a sufficiently long time so that the local ecosystem reaches its equilibrium. Then we compute the biogenic fluxes between the surface ocean and the atmosphere, and between the surface ocean and the deep ocean, and feed them into the planetary model. The planetary model is then used to resolve the system's geochemical and climate dynamics, but only for a sufficiently small change in the planetary state (1% to 1‰ variation of the state variable that changes the most) so that this change would alter the biological equilibrium only marginally. The new biological equilibrium is then computed, and the process is re-iterated. This iterative process approximates a continuous coupling between the microbial community and its planetary environment.

**Monte-Carlo simulations**

We explore the parameter range by stochastically varying the parameters that determine the maximum metabolic rate, the rate of energy consumption for maintenance, the metabolic half-saturation constant, the decay and mortality rates, and the maximum division rate. We also vary the values of the parameters that shape the dependencies on size, temperature, and cellular biomass (see Supplementary Table 4). Parameter values are picked uniformly or log-uniformly (to avoid making prior assumptions on the likelihood of specific regions of the parameter space) within ranges constrained by empirical data from the literature, or large enough to cover plausible empirical variation. Three thousand simulations were run for the MG ecosystem, one thousand for the AG+AT ecosystem, and one thousand for the MG+AG+AT ecosystem, under four different values of volcanic outgassing: $\phi_{Volc}(H_2)$ = 2 $10^{9.5}$, 2 $10^{10}$, 2 $10^{10.5}$, and 2 $10^{11}$ molecules $cm^{-2}$ $s^{-1}$, hence a total of 20,000 simulations. Each simulation was run to equilibrium.

The results can be divided into two subsets of roughly equivalent size, regardless of the ecosystems and volcanic activity considered: those with biological activity and those without (Supplementary Figure 1A and B). We then verified that the number of simulations run for each of the considered scenarios was sufficient for the resulting distribution of equilibrium states to have converged. We evaluated convergence by subsampling, for each scenario, the subset of 'biologically viable' simulations with increasing sample size, and computing the average equilibrium biomass and $CH_4$ biogenic emission of the subsamples. The results are shown in Supplementary Figure 2 for the three scenarios and for $\phi_{Volc}(H_2)$= 2 $10^{10.5}$ molecules $cm^{-2}$ $s^{-1}$.

**Data availability**

The data used to produce all the results presented in this study are available upon reasonable request to B.S..

**Code availability**

The codes used in this study to produce the data analysed are available on a git repository upon reasonable request to B.S..

**References**

1. Nisbet, E. G., and N. H. Sleep. "The habitat and nature of early life." *Nature* 409.6823 (2001): 1083.

2. Martin, William F., and Filipa L. Sousa. "Early microbial evolution: the age of anaerobes." *Cold Spring Harbor perspectives in biology* 8.2 (2016): a018127.

3. Pearce, Ben KD, et al. "Constraining the time interval for the origin of life on Earth." *Astrobiology* 18.3 (2018): 343-364.

4. Charnay, Benjamin, et al. "A warm or a cold early Earth? New insights from a 3-D climate-carbon model." *Earth and Planetary Science Letters* 474 (2017): 97-109.

5. Krissansen-Totton, Joshua, et al. "Constraining the climate and ocean pH of the early Earth with a geological carbon cycle model." *Proceedings of the National Academy of Sciences* (2018): 201721296.

6. Catling, David C., and James F. Kasting. *Atmospheric evolution on inhabited and lifeless worlds*. Cambridge University Press, 2017.

7. Battistuzzi, Fabia U., et al. "A genomic timescale of prokaryote evolution: insights into the origin of methanogenesis, phototrophy, and the colonization of land." *BMC evolutionary biology* 4.1 (2004): 44.

8. Ozaki, Kazumi, et al. "Effects of primitive photosynthesis on Earth's early climate system." *Nature Geoscience* 11.1 (2018): 55.

9. Weiss, Madeline C., et al. "The physiology and habitat of the last universal common ancestor." *Nature Microbiology* 1.9 (2016): 16116.

10. Marin, Julie, et al. "The timetree of prokaryotes: new insights into their evolution and speciation." *Molecular biology and evolution* 34.2 (2016): 437-446.

11. Havig, Jeff R., et al. "Sulfur and carbon isotopic evidence for metabolic pathway evolution and a four-stepped Earth system progression across the Archean and Paleoproterozoic." *Earth-Science Reviews* 174 (2017): 1-21.

12. Kasting, James F., et al. "A coupled ecosystem-climate model for predicting the methane concentration in the Archean atmosphere." *Origins of Life and Evolution of the Biosphere* 31.3 (2001): 271-285.

13. Kharecha, P., J. Kasting, and J. Siefert. "A coupled atmosphere–ecosystem model of the early Archean Earth." *Geobiology* 3.2 (2005): 53-76.

14. Ward, Lewis M., Birger Rasmussen, and Woodward W. Fischer. "Primary productivity was limited by electron donors prior to the advent of oxygenic photosynthesis." *Journal of Geophysical Research: Biogeosciences* 124.2 (2019): 211-226.

15. Charnay, Benjamin, et al. "Exploring the faint young Sun problem and the possible climates of the Archean Earth with a 3-D GCM." *Journal of Geophysical Research: Atmospheres* 118.18 (2013): 10-414.

16. Lefèvre, Franck, et al. "Three-dimensional modeling of ozone on Mars." *Journal of Geophysical Research: Planets* 109.E7 (2004).

17. Kleerebezem, Robbert, and Mark CM Van Loosdrecht. "A generalized method for thermodynamic state analysis of environmental systems." *Critical Reviews in Environmental Science and Technology* 40.1 (2010): 1-54.

18. González-Cabaleiro, Rebeca, Juan M. Lema, and Jorge Rodríguez. "Metabolic energy-based modelling explains product yielding in anaerobic mixed culture fermentations." *PLoS One* 10.5 (2015): e0126739.

19. Tijhuis, L., Mark CM Van Loosdrecht, and J. J. Heijnen. "A thermodynamically based correlation for maintenance Gibbs energy requirements in aerobic and anaerobic chemotrophic growth." *Biotechnology and bioengineering* 42.4 (1993): 509-519.

20. Gillooly, James F., et al. "Effects of size and temperature on metabolic rate." *science* 293.5538 (2001): 2248-2251.

21. Litchman, Elena, et al. "The role of functional traits and trade-offs in structuring phytoplankton communities: scaling from cellular to ecosystem level." *Ecology letters* 10.12 (2007): 1170-1181.

22. Ward, Ben A., et al. "The size dependence of phytoplankton growth rates: a trade-off between nutrient uptake and metabolism." *The American Naturalist* 189.2 (2017): 170-177.

23. Aksnes, D. L., and J. K. Egge. "A theoretical model for nutrient uptake in phytoplankton." Marine ecology progress series. Oldendorf 70.1 (1991): 65-72.

24. Arney, Giada, et al. "The pale orange dot: the spectrum and habitability of hazy Archean Earth." *Astrobiology* 16.11 (2016): 873-899.

25. Domagal-Goldman, Shawn D., et al. "Organic haze, glaciations and multiple sulfur isotopes in the Mid-Archean Era." *Earth and Planetary Science Letters* 269.1-2 (2008): 29-40.

26. Prentice, Iain Colin, et al. "The carbon cycle and atmospheric carbon dioxide." Cambridge University Press, 2001. 185-237.

27. Moore, Eli K., et al. "Metal availability and the expanding network of microbial metabolisms in the Archaean eon." *Nature Geoscience* 10.9 (2017): 629-636.

28. Catling, D. C., M. W. Claire, and K. J. Zahnle. "Anaerobic methanotrophy and the rise of atmospheric oxygen." *Philosophical Transactions of the Royal Society A: Mathematical, Physical and Engineering Sciences* 365.1856 (2007): 1867-1888.

29. Wong, Michael L., et al. "Nitrogen oxides in early Earth's atmosphere as electron acceptors for life's emergence." Astrobiology 17.10 (2017): 975-983.

30. Ono, Shuhei, et al. "New insights into Archean sulfur cycle from mass-independent sulfur isotope records from the Hamersley Basin, Australia." Earth and Planetary Science Letters 213.1-2 (2003): 15-30.

31. Habicht, Kirsten S., et al. "Calibration of sulfate levels in the Archean ocean." *Science* 298.5602 (2002): 2372-2374.

32. Crowe, Sean A., et al. "Sulfate was a trace constituent of Archean seawater." *Science* 346.6210 (2014): 735-739.

33. Kasting, James F. "Methane and climate during the Precambrian era." *Precambrian Research* 137.3-4 (2005): 119-129.

34. Kasting, James F., and Shuhei Ono. "Palaeoclimates: the first two billion years." Philosophical Transactions of the Royal Society B: Biological Sciences 361.1470 (2006): 917-929.

35. Krissansen-Totton, et al. "A statistical analysis of the carbon isotope record from the Archean to Phanerozoic and implications for the rise of oxygen." American Journal of Science 315.4 (2015): 275-316.

36. Knittel, Katrin, and Antje Boetius. "Anaerobic oxidation of methane: progress with an unknown process." Annual review of microbiology 63 (2009): 311-334.

37. Nitschke, Wolfgang, and Michael J. Russell. "Beating the acetyl coenzyme A-pathway to the origin of life." Philosophical Transactions of the Royal Society B: Biological Sciences 368.1622 (2013): 20120258.

38. Sleep, Norman H., and Kevin Zahnle. "Carbon dioxide cycling and implications for climate on ancient Earth." Journal of Geophysical Research: Planets 106.E1 (2001): 1373-1399.

39. Laakso, Thomas A., and Daniel P. Schrag. "Methane in the Precambrian atmosphere." Earth and Planetary Science Letters 522 (2019): 48-54.

40. Lyons, Timothy W., Christopher T. Reinhard, and Noah J. Planavsky. "The rise of oxygen in Earth's early ocean and atmosphere." Nature 506.7488 (2014): 307.

41. Kopp, Robert E., et al. "The Paleoproterozoic snowball Earth: a climate disaster triggered by the evolution of oxygenic photosynthesis." Proceedings of the National Academy of Sciences 102.32 (2005): 11131-11136.

42. Krissansen-Totton, Joshua, Stephanie Olson, and David C. Catling. "Disequilibrium biosignatures over Earth history and implications for detecting exoplanet life." *Science advances* 4.1 (2018): eaao5747.

43. González-Cabaleiro, Rebeca, et al. "Microbial catabolic activities are naturally selected by metabolic energy harvest rate." *The ISME journal* 9.12 (2015): 2630.

44. Tilman, David. *Resource competition and community structure*. Princeton university press, 1982.

45. Mylius, Sido D., and Odo Diekmann. "On evolutionarily stable life histories, optimization and the need to be specific about density dependence." *Oikos* (1995): 218-224.

46. Metz, J. A. J., S. D. Mylius, and O. Diekmann. "When does evolution optimise?" (2008).

47. Wordsworth, R., et al. "Modelling past Mars Climates and water cycle with a thicker CO2 atmosphere." *Mars Atmosphere: Modelling and Observation* (2011): 447-448.

48. Leconte, Jérémy, et al. "Increased insolation threshold for runaway greenhouse processes on Earth-like planets." *Nature* 504.7479 (2013): 268.

49. Codron, Francis. "Ekman heat transport for slab oceans." *Climate Dynamics* 38.1-2 (2012): 379-389.

50. Lefèvre, F., S. Lebonnois, and F. Forget. "A three-dimensional photochemical-transport model of the martian atmosphere." *Sixth International Conference on Mars*. 2003.

51. Arney, Giada, et al. "Hazy Archean Earth as an Analog for Hazy Earthlike Exoplanets." *American Astronomical Society Meeting Abstracts# 225*. Vol. 225. 2015.

52. Pavlov, Alexander A., Lisa L. Brown, and James F. Kasting. "UV shielding of NH3 and O2 by organic hazes in the Archean atmosphere." *Journal of Geophysical Research: Planets* 106.E10 (2001): 23267-23287.

53. Zahnle, K., M. Claire, and D. Catling. "The loss of mass-independent fractionation in sulfur due to a Palaeoproterozoic collapse of atmospheric methane." *Geobiology* 4.4 (2006): 271-283.

54. Claire, Mark W., et al. "The evolution of solar flux from 0.1 nm to 160 μm: quantitative estimates for planetary studies." *The Astrophysical Journal* 757.1 (2012): 95.

55. Hu, Renyu, Luke Peterson, and Eric T. Wolf. "O2-and CO-rich Atmospheres for Potentially Habitable Environments on TRAPPIST-1 Planets." *The Astrophysical Journal* 888.2 (2020): 122.

**Acknowledgements** We thank Daniel Apai, Alex Bixel, Dillon Demo, Zach Grochau-Wright, François Guyot, Betul Kacar, James Kasting, Charles Lineweaver, Scot Rafkin and Alexander Sotto for discussion; and three anonymous reviewers for their insightful comments and helpful suggestions. This work was supported by Paris Sciences & Lettres University (IRIS OCAV and PSL-University of Arizona Mobility Program). B.S. acknowledges support from PSL IRIS *Origins and conditions for the emergence of life* (OCAV) program. R.F. acknowledges support from the United States National Science Foundation, *Dimensions of Biodiversity* program (DEB-1831493).

**Author contributions** R.F. and S.M. conceived the study. R.F. designed the ecosystem model. A.A. and B.S. refined the model and developed the code. B.C. provided the climate and carbon cycle modules. B.S. and S.M. designed the Monte-Carlo approach and ran the simulations. B.S. analyzed the results and wrote the first version of the manuscript. All authors finalized the paper.

**Competing interests** The authors declare no competing interests.

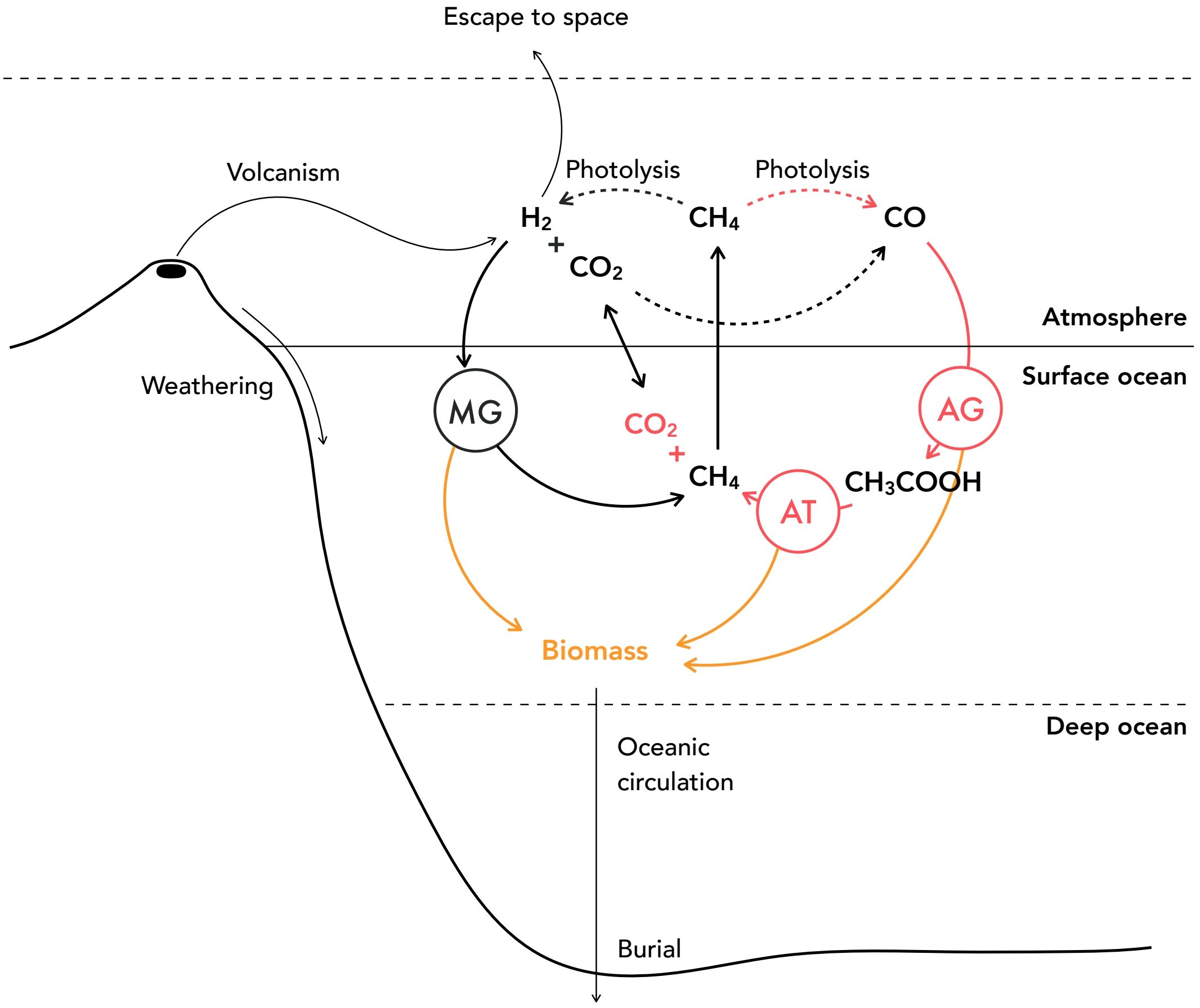

**Figure 1. Primitive methanogenic ecosystems.** The ecosystem model resolves population abundance (total biomass, yellow arrows) of microbial $H_2$-based methanogens (MG), CO-based acetogens (AG) and methanogenic acetotrophs (AT), along with $CH_4$, CO, $CO_2$, and $H_2$ oceanic concentrations and atmospheric mixing ratios. Fluxes directly involved in the MG ecosystem function are indicated with black arrows. Fluxes additionally involved in the AG+AT ecosystem function are indicated in red. Key photochemical reactions are indicated with dotted arrows. The primary source of reducing power ($H_2$) is volcanic outgassing. Fluxes across the ocean surface are governed by a stagnant boundary layer model. Rates of $H_2$ escape to space and dead biomass burial in deep sediments are constant. Sulfate-based methanotrophs are not represented. See Methods for further details.

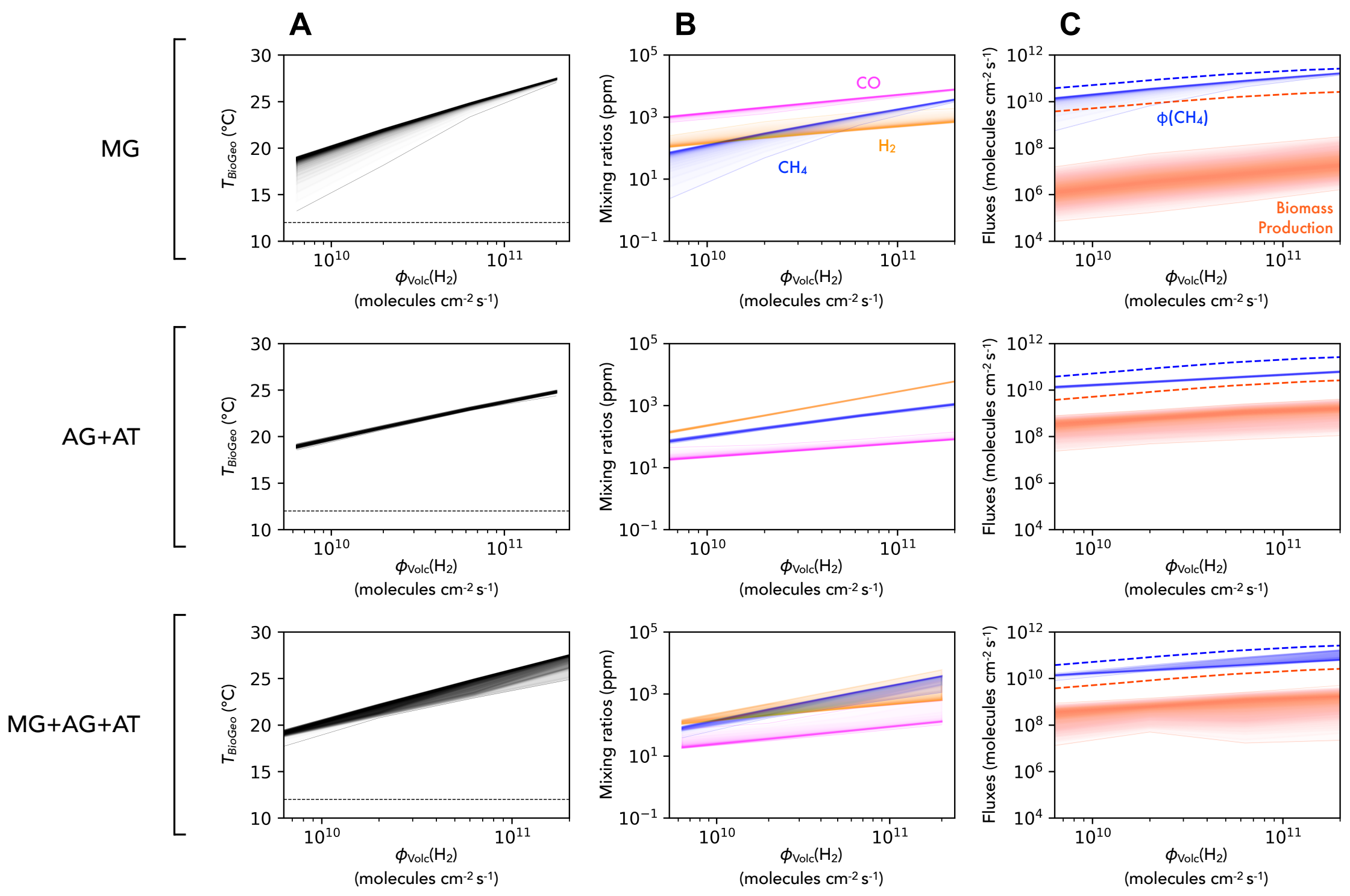

**Figure 2. Short-term biological feedback to the atmosphere and climate.** Effects are computed as a function of the H2 volcanic outgassing, for each ecosystem composition. **MG** indicates $H_2$-based methanogens (MG). **AG+AT** indicates CO-based acetogens and methanogenic acetotrophs consortia. **MG+AG+AT** indicates co-occurring methanogens, acetogens, and acetotrophs. (A)*,* Global surface temperature at ecosystem-climate equilibrium. The dotted line indicates the initial abiotic surface temperature, $T_{Geo}$ = 12 °C. (B)*,* Atmospheric composition at ecosystem-climate equilibrium. (C)*,* Biogenic fluxes at ecosystem-climate equilibrium: $CH_4$ production and carbon fixation in biomass (in molecules of C $cm^{-2}$ $s^{-1}$). Enveloppes represent probability distributions from Monte-Carlo simulations across the biological parameter space, with each layer indicating output frequency ranging from 90% to 51%. Predictions from ref[13] are also shown (dashed) for comparison. See Methods and Supplementary Tables 3 and 4 for parameter values.

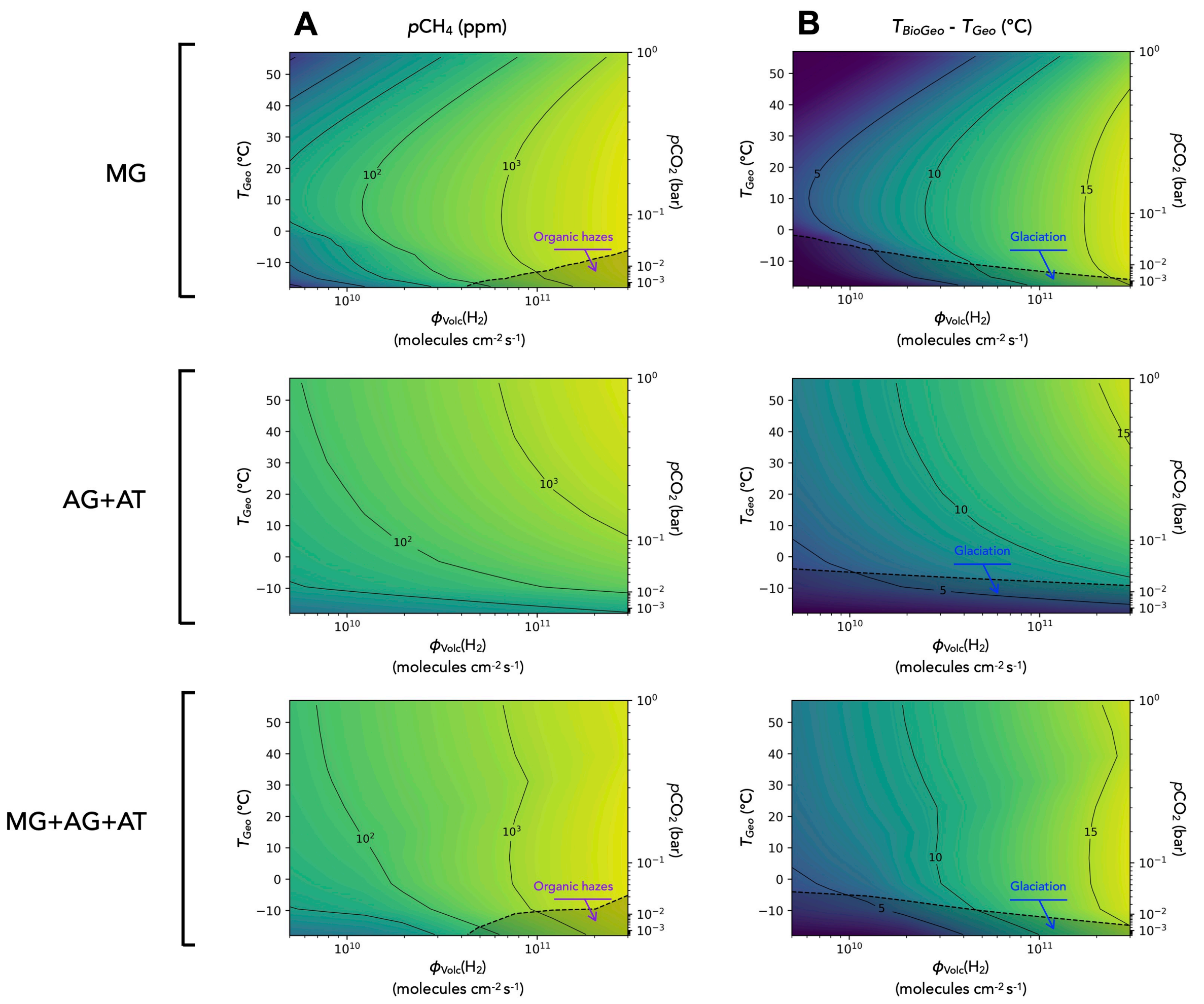

**Figure 3. Short-term biological feedback on the atmosphere and climate.** Effects are computed as a function of the $H_2$ volcanic outgassing and abiotic surface temperature ($T_{Geo}$), for each ecosystem composition**. MG** indicates $H_2$-based methanogens. **AG+AT** indicates CO-based acetogens and methanogenic acetotrophs consortia. **MG+AG+AT** indicates co-occurring methanogens, acetogens, and acetotrophs. $T_{Geo}$ is varied by changing $pCO_2$ in the climate model. (A), Atmospheric $pCH_4$ at ecosystem-climate equilibrium. Shaded areas indicate conditions for organic haze formation. (B), Temperature differential between $T_{Geo}$ and the global surface temperature reached at ecosystem-climate equilibrium, $T_{BioGeo}$. Shaded areas indicate conditions leading to organic haze formation (A) and glaciation (B). Other parameters are set at their default values (Supplementary Tables 2 and 3).

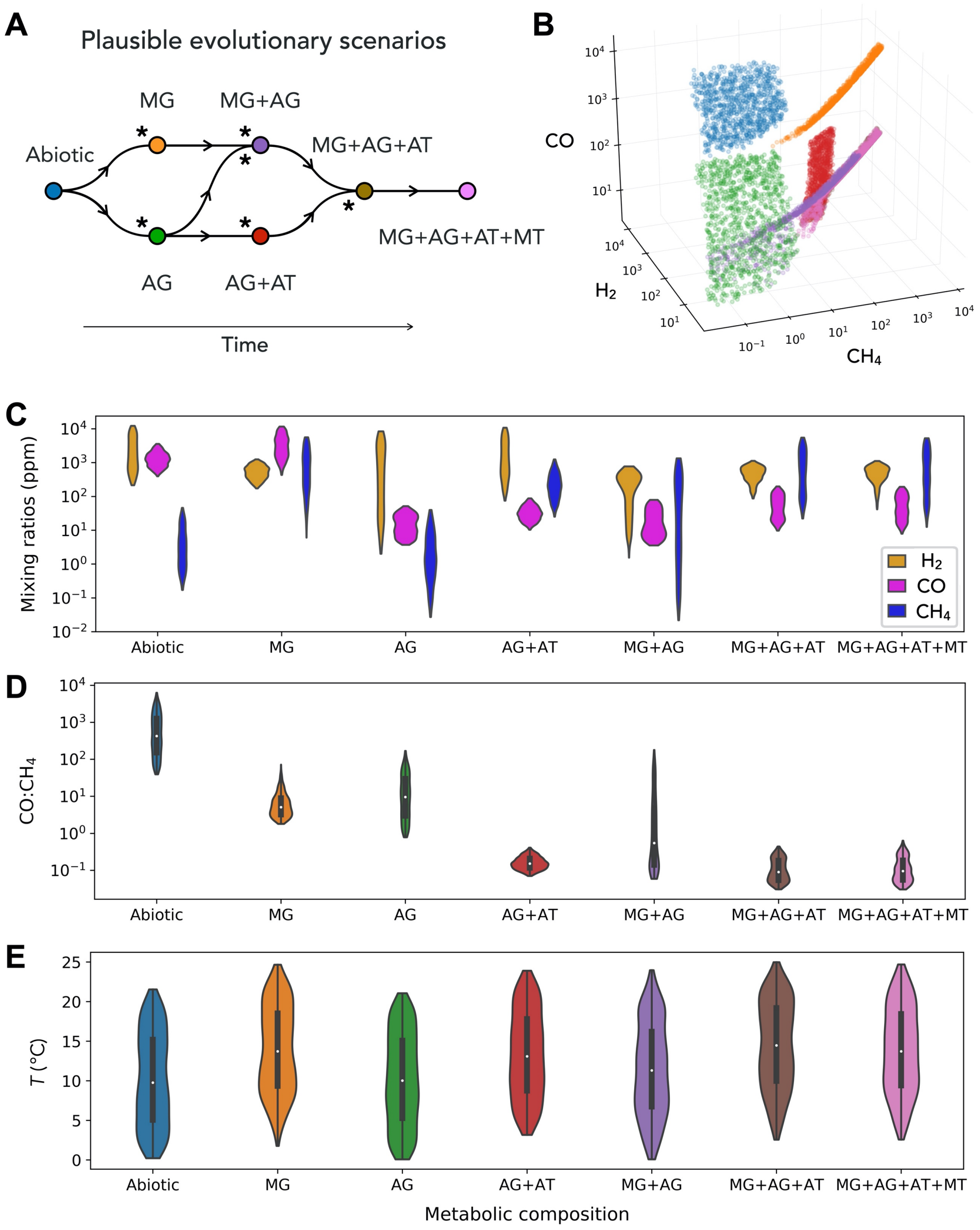

**Figure 4. Equilibrium state of the planet as the biosphere diversifies.** (**A**) Plausible evolutionary sequences of metabolic innovation. Asterisks denote transitions that are very likely to cause a significant change in the atmospheric composition. (**B**) Scatterplot of the atmospheric compositions at equilibrium in CO, $CH_4$ and $H_2$, color-coded by the corresponding biosphere composition (1,000 simulations for each ecosystem). (**C**) Corresponding distributions. (**D**) Distribution of the CO:$CH_4$ ratio for each scenario. (**E**) Distribution of the surface temperature in each scenario. The white dots in **D** and **E** represent the median of the distributions, the thick gray lines the interquartile range, and thin gray lines the rest of the distribution.

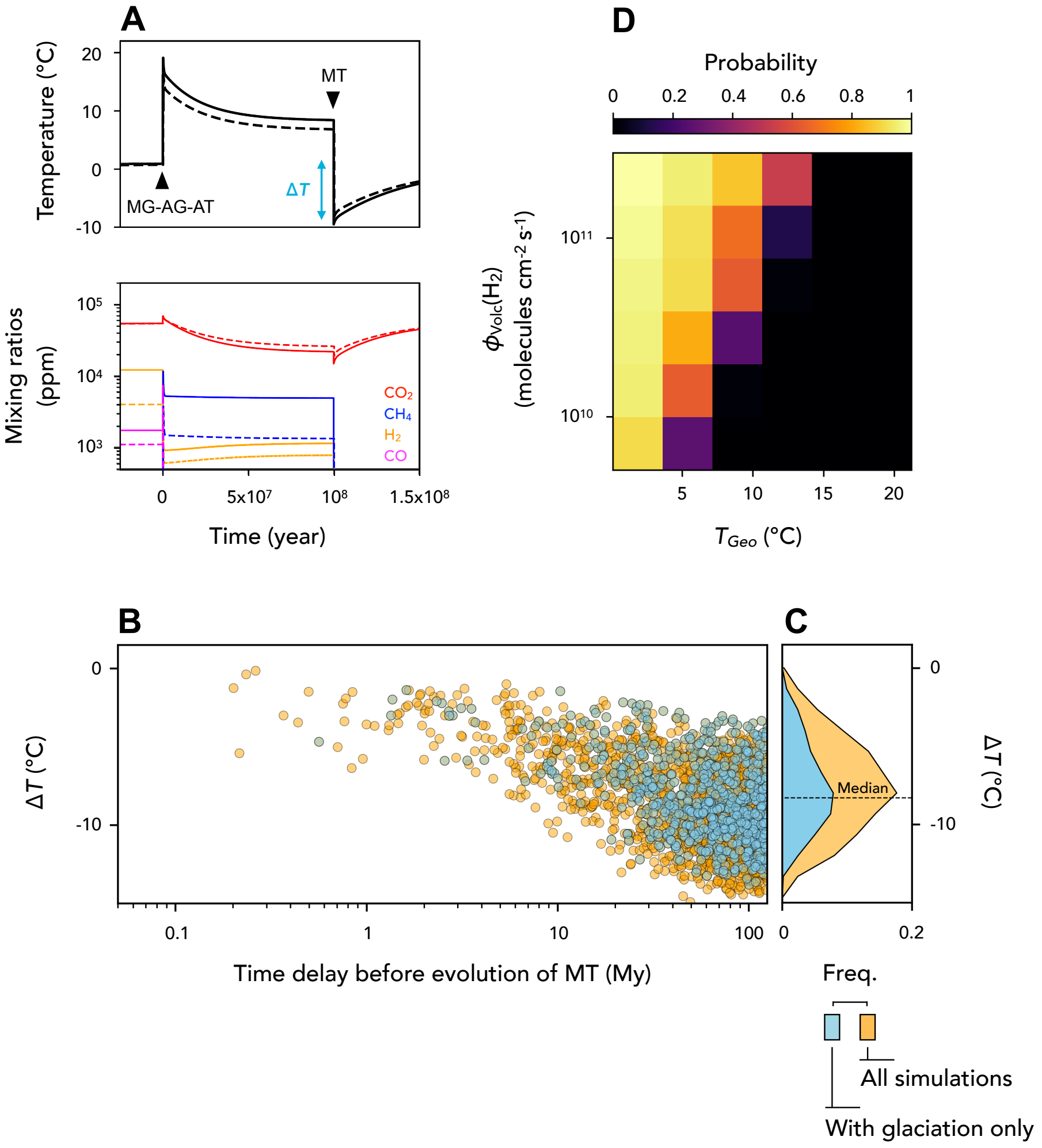

**Figure 5. Climate destabilization by evolutionary metabolic innovation.** (**A**) In this example, sulfur-based methanotrophy (MT) evolves 100 million years after MG+AG+AT, i.e. after equilibration of the methanogenic biosphere (MG+AG+AT) with the atmosphere and climate mediated by the carbon cycle, with $T_{Geo}$ = 2 °C, $\varphi_{volc}(H_2) = 3\ 10^{11}$ (plain lines) and $1\ 10^{11}$ molecules $s^{-1}$ $cm^{-2}$ (dotted lines). *Top*, Change in surface temperature. *Bottom*, Change in atmospheric composition. (**B to D**) Distribution of outcomes across a range of abiotic temperature $T_{Geo}$, $H_2$ volcanic flux, and evolution time of MT (2,000 randomly chosen combinations). (B), Amplitude of global cooling, $\Delta T$, with respect to the evolution time of MT. (C), Frequency distribution of all temperature changes $\Delta T$ (blue) and of temperature changes conditional on glaciation outcome (yellow). (D) Estimated probability of glaciation as a consequence of MT evolution, given the abiotic temperature $T_{Geo}$ and $H_2$ volcanic flux. Other parameters are set to their default values (Supplementary Tables 2 and 3).

**Table 1. Abiotic inputs of $CO_2$, $H_2$, $CH_4$ and $H_2SO_4$**

| **Abiotic inputs** | **Range (in molecules $cm^{-2}$ $s^{-1}$)** | **Reference** |
|---|---|---|
| Volcanic output of $CO_2$ | $1.2\ 10^{10} – 4\ 10^{10}$ (to obtain $T_{Geo}$ ranging from 0 to 20°C ) | ref[5] |
| Volcanic output of $H_2$ | $5\ 10^{9} – 3\ 10^{11}$ | ref[12] |
| Serpentinization rate of production of $CH_4$ | $3.7\ 10^{8} – 3.7\ 10^{9}$ | Ref[33,42] |
| Deposition rate of $H_2SO_4$ | $10^{7} – 10^{9}$ | ref[30] |

Supplementary Information for

“Co-evolution of primitive methane-cycling ecosystems and early Earth’s atmosphere and climate”

by Boris Sauterey et al.

Correspondence to: boris.sauterey@ens.fr

## Supplementary Results and Discussion

Infuence of biological parameters on ecosystem viability

By comparing the distribution of parameter values from the subset of simulations with persistent biological activity to the distribution of all parameter values, we can delineate the region in parameter space that corresponds to ecosystem viability (Supplementary Figure 1C). We find that viability is significantly conditioned by high values of $b_q$, low values of basal mortality $m$, low values of $b_E$, and high values of maximum division rate $r_{max}$, with a predominant effect of $b_q$ (Supplementary Figure 1D and E).

Next, we use the subset of ecologically viable simulations to examine how model parameters influence the equilibrium biomass and biogenic methane flux, $\phi_{Bio}(CH_4)$. We find that the level of biogenic methane emission is positively related to the maximum metabolic rate through parameter $b_q$ and negatively related to the maintenance cost through $b_E$, while biomass production negatively correlates with both $b_E$ and $b_q$(statistical analysis not shown).

Finally, we compare the distribution of outputs in the subset of ecologically viable simulations to the default parameterization outcome (Fig. 1). The distribution is relatively narrow (95% interval envelope is about one order of magnitude wide), highlighting the fact that the model is more strongly constrained by its structure than parameterization. In most scenarios, default parameter values yield results that are close to the median of the subset of ecologically viable simulations (Fig. 2). With the MG ecosystem, the results of the default parametrization are within the 95% confidence intervals, but close to the boundaries (lower limit for methane emission, upper limit for biomass; data not shown). This greater sensitivity to parameters is due to the global redox equilibrium being strongly impacted by the metabolic rate of methanogens; this is in contrast to the other ecological scenarios where the global redox equilibrium is determined chiefly by photochemical processes. The default value of $q_{max}$ is near the lower end of the viability range for that parameter, so most of the ecologically viable simulations correspond to higher $q_{max}$, hence larger $CH_4$ emission at equilibrium and lower equilibrium biomass. This is because the redox state of the system is closer to its thermodynamic equilibrium, and therefore metabolism is less efficient. Interestingly, exploring higher values of $q_{max}$ is equivalent to releasing kinetic constraints on biology. This explains why our predictions of $CH_4$ emission then get closer to ref[12].

Global redox balance of the planet

By tracking the global hydrogen budget of the atmosphere, computed as $f(H_2) + 4\ f(CH_4) + f(CO)$ following ref[6], we check the evolution of the atmospheric global redox budget in the simulations presented in Fig 4. We find that in most cases the atmospheric redox budget is very similar whether the planet is populated by a primitive biosphere or not (Supplementary Figure 11). The only two exceptions are the ecological scenarios in which AG is present in the biosphere in the absence of AT. When this is the case, some of the redox budget of the atmosphere is transferred to the ocean in the form of acetate.

The conservation of a steady atmospheric redox budget from a lifeless to a living Earth highlights that the biomass production of primitive biospheres was so low that it did not constitute a significant sink of $H_2$ relative to atmospheric escape (the main sink of reduced species during the Archean).

## Ecological feedback of methanogenesis on climate warming and resilience

Supplementary Figure 3 shows the atmospheric and climatic impact of the biosphere as a function of $H_2$ volcanic outgassing, $\phi_{Volc}(H_2)$, and the abiotic temperature, $T_{Geo}$. Here $T_{Geo}$ varies independently of $pCO_2$ due to external factors such as solar activity. Two values of $pCO_2$ are tested: 2500 ppm (Supplementary Figure 3A) to allow comparison with ref[12], and $10^5$ ppm (Supplementary Figure 3B) for comparison with Fig. 1 and 3 (the climate model predicts $10^5$ ppm $CO_2$ to set $T_{Geo}$ at 12 °C). Qualitatively, the results are similar to those reported in Fig. 3 where temperature is set by $pCO_2$. Quantitatively, with the MG ecosystem the effect of temperature variation on biological activity is even stronger when $T_{Geo}$ and $pCO_2$ are independent. This is because the negative effect of higher temperature on thermodynamics is partially offset if $pCO_2$ is concomitantly higher. As a consequence, hydrogenotrophic methanogenesis is expected to have an even greater effect on climate when temperature varies independently of $pCO_2$. In contrast, climate warming by AG+AT metabolisms is weak, irrespective of $H_2$ outgassing and abiotic temperature when the latter varies independently of $pCO_2$. Warming will occur, however, in an AG+AT ecosystem in which MG evolves, the effect being as strong as in the MG-only ecosystem (Supplementary Figure 3A).

Climate resilience to variation of $pCO_2$ is shown in Supplementary Figure 4. With the MG ecosystem, the ecological feedback to the atmosphere has a buffering effect on temperature variation above *ca.* 5 °C (Supplementary Figure 4A) and an amplifying effect below 5 °C (Supplementary Figure 4B). With the AG+AT ecosystems, the amplification effect prevails irrespective of the temperature range (Supplementary Figure 4C and D). Once MG, AG and AT have all evolved, we can however conclude from Fig. 3 that the ecosystem has almost no effect on the resilience of the climate.

## Atmospheric and climatic impact of methanotrophy

$CH_4$ emissions by MG and AG+AT metabolisms create conditions favorable for the evolution of methanotrophy (MT). The evolutionary rate has a critical influence on the MT environmental feedback. Supplementary Figure 6 shows the environmental impact of fast-evolving MT that arises on a $10^3$ years timescale after the establishment of MG and/or AG+AT metabolisms. In this case, MT evolution takes place under atmospheric and climatic conditions set by the atmosphere-ecosystem equilibrium of MG and/or AG+AT (Figs. 2 and 3, Supplementary Figure 3). Irrespective of $H_2$ volcanic outgassing rate and abiotic temperature (Supplementary Figure 6A), the environmental effect of MT is to consume most of the atmospheric $CH_4$ produced by methanogens (Supplementary Figure 6C), driving the surface temperature close to its abiotic value, $T_{Geo}$ (Supplementary Figure 6B). The timescale over which this happens is very short, of the order of $10^3$ years (Supplementary Figure 6B and C). The outcome is a new atmosphere-ecosystem equilibrium at which all metabolisms coexist, under a methane-poor atmosphere resulting in a cool climate.

The previous scenario will hold provided the evolutionary timescale is much shorter than the timescale of the carbon cycle. If the timescale of MT evolution is of the order of the C cycle characteristic time ($10^7$ yrs), or longer, then the environmental impact of MT evolution will depend on the long-term effect that the carbon cycle has on the environment inhabited by MG and/or AG+AT ecosystems (Supplementary Figure 12). As explained in the main text, the carbon cycle response to the evolution of methanogenesis leaves the biogenic outflux of $CH_4$ relatively unaltered. However, the equilibrium temperature, $T_{BioGeo}$, is much lower than at the short-term equilibrium shown in Fig. 3, in the absence of carbon cycle feedback. Under such conditions, the evolution of methanotrophy drives both $pCH_4$ and $pCO_2$ down (Fig. 5A), causing dramatic climate cooling and putting the planet at high risk of global glaciation (Fig. 5B and C). Supplementary Figure 8C shows that relatively low abiotic temperature combined with a high rate of $H_2$ volcanic outgassing favors the global glaciation outcome.

Supplementary Figures

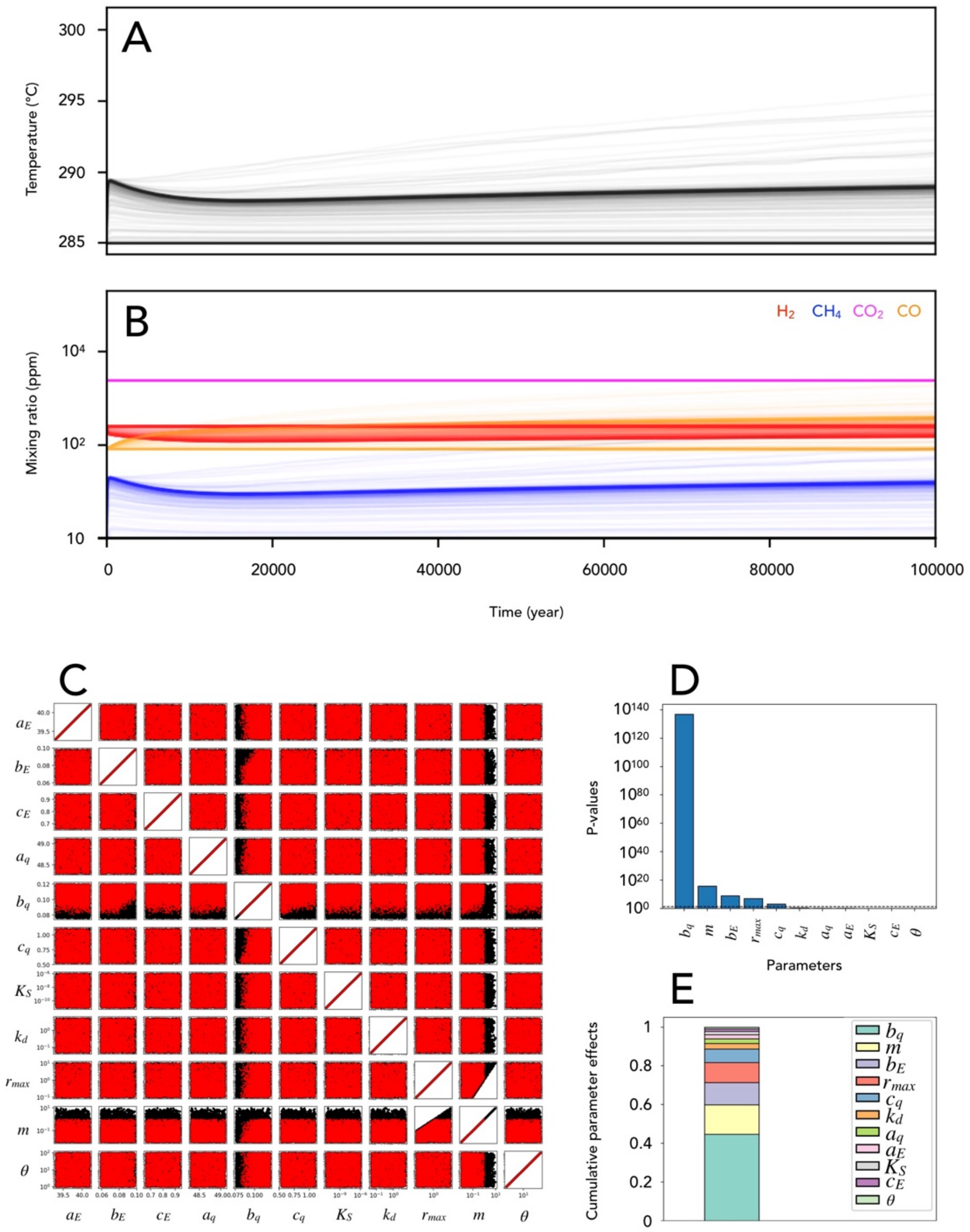

**Supplementary Figure 1.**

**Outputs of 5,000 Monte Carlo simulations for the MG ecosystem.** (**A**) Surface temperature. (**B**) Atmospheric composition. (**C**) Ecosystem viability across the parameter space. Red dots indicate simulations in which the ecosystem is viable; other simulations are indicated in black. A discrepancy between the distributions of black and red dots for a given parameter indicates that viability is favored by a specific range in that parameter's value. The significance and strength of each parameter's influence on MG ecosystem viability is given in panels **D** and **E**, respectively. Volcanic $H_2$ outgassing $\phi_{Volc}(H_2)$ is set to 2 $10^{9.5}$, other parameters are set to their default values (Supplementary Tables 2 and 3).

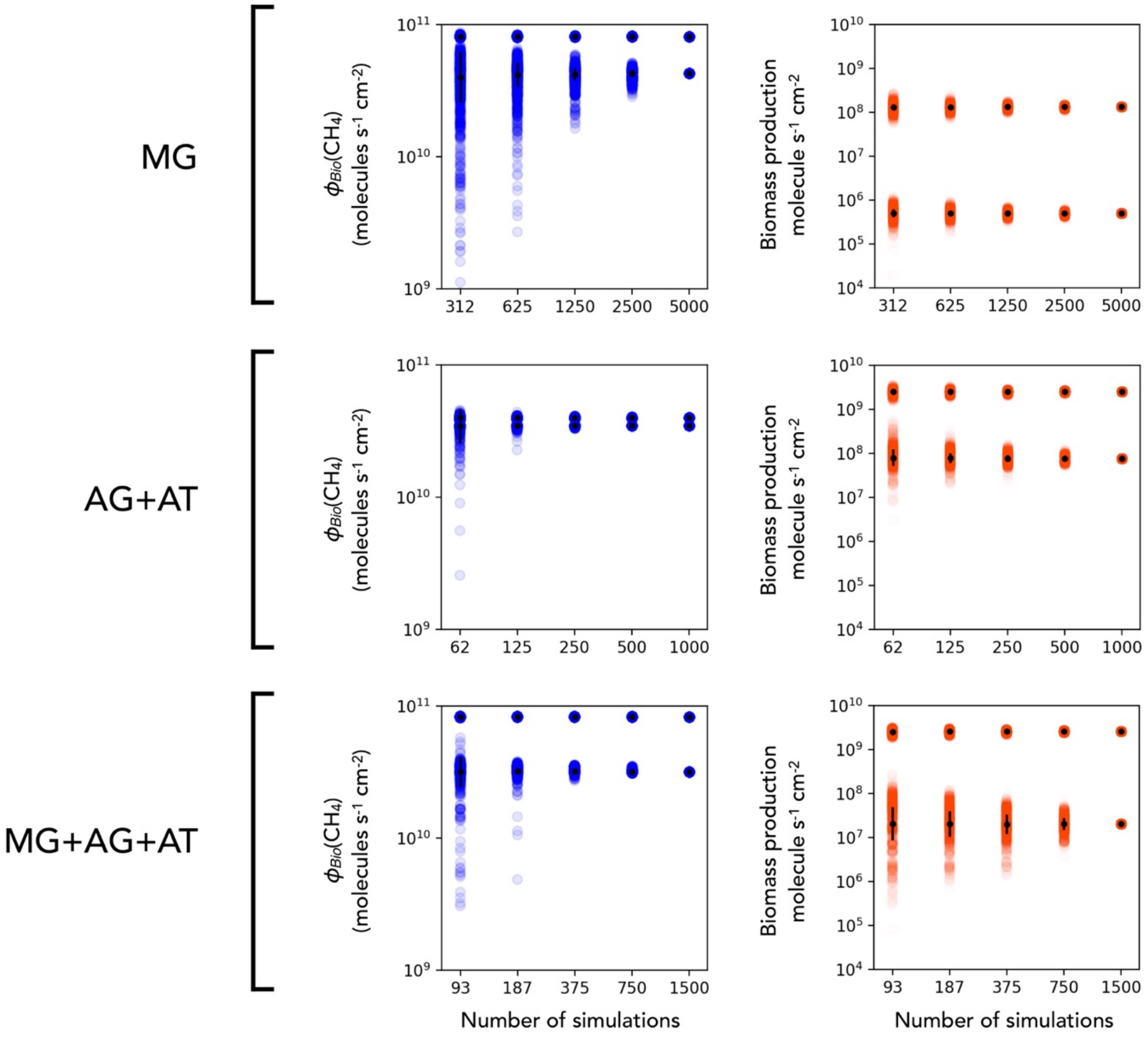

**Supplementary Figure 2.**

**Influence of the number of simulations on the higher and lower boundaries of biomass production and biogenic emission of CH4, $\phi_{Bio}(CH_4)$.** The numbers of simulations represent 1/16, ⅛, ¼, ½ and the full set of the whole simulations ensemble. For each simulation set size, a thousand subsamples of that size are bootstrapped, for which the average value (dot) and standard deviation (error bar) are calculated. Noticingly, for all three types of ecosystems, the lower boundary is largely underestimated when the number of simulations is too low, and converge toward its actual value as the number of simulations increases. $H_2$ volcanic outgassing is fixed atof 2 $10^{10.5}$ molecules cm$^{-2}$ s$^{-1}$, other parameters are set to their default values (Supplementary Tables 2 and 3).

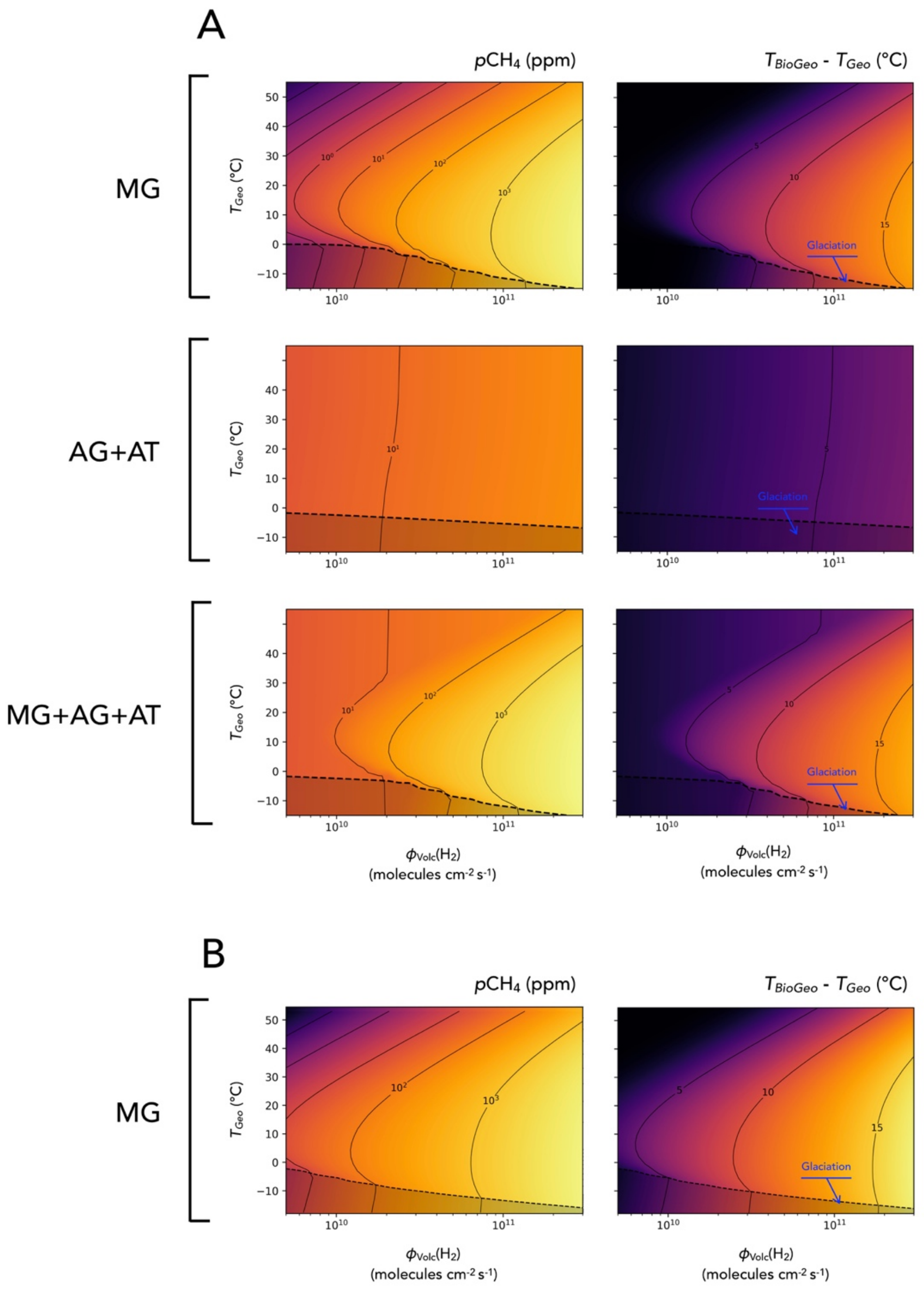

**Supplementary Figure 3.**
**Biogeochemical response of the early Archean Earth to changes in $H_2$ volcanic outgassing, abiotic temperature ($T_{Geo}$) and ecosystem composition.** Here $T_{Geo}$ is varied independently of $pCO_2$ in the climate model. (A) $pCO_2$ = 2500 ppm. (B) $pCO_2 = 10^5$ ppm. *Left*, Atmospheric $pCH_4$ at ecosystem-climate equilibrium. Shaded areas indicate conditions for organic haze formation. *Right*, Temperature differential between $T_{Geo}$ and the global surface temperature reached at ecosystem-climate equilibrium, $T_{BioGeo}$ . Shaded areas indicate conditions leading to glaciation. Other parameters are fixed to their default values (Supplementary Tables S2 and S3).

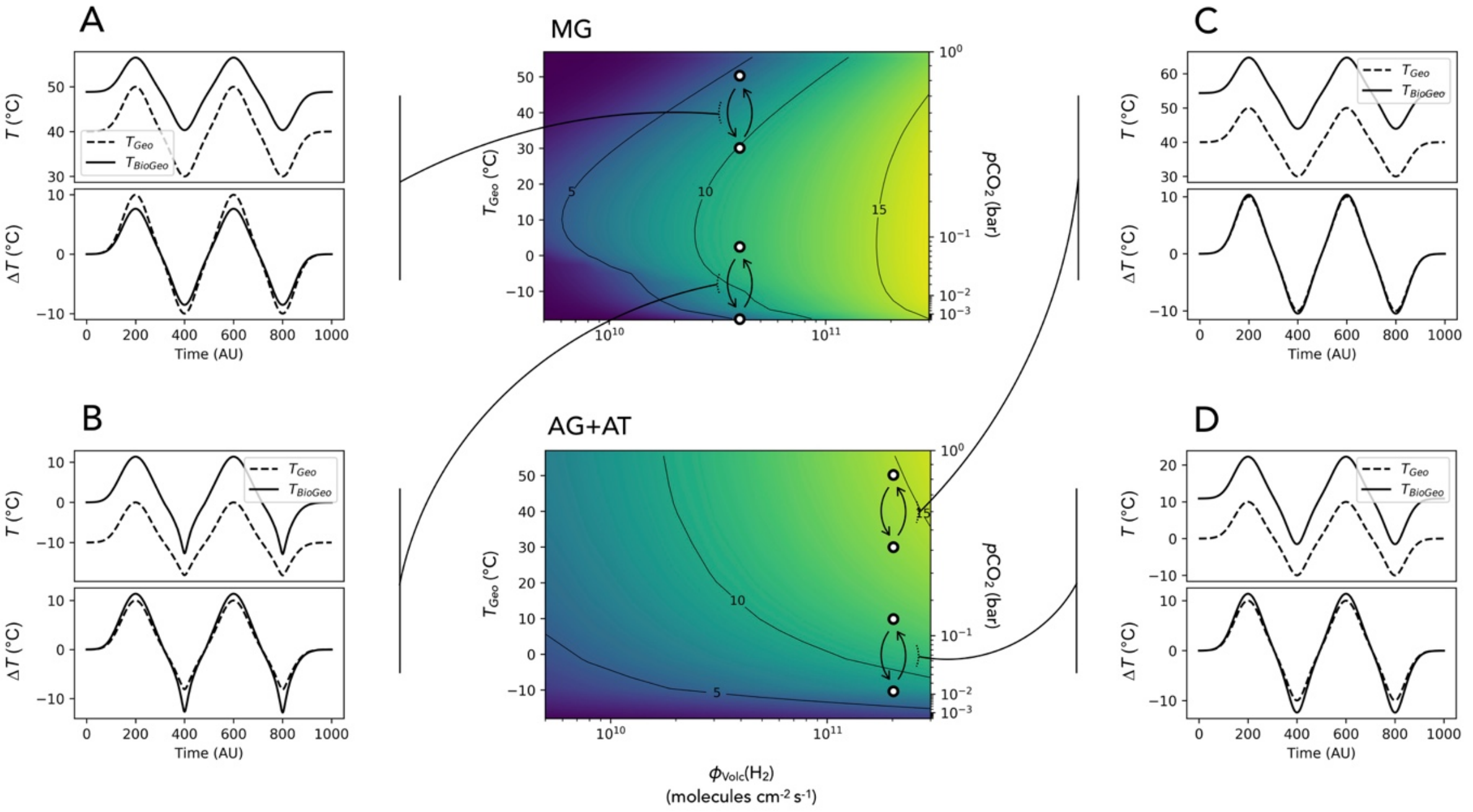

**Supplementary Figure S4.**

**Climate resilience in response to $p\mathrm{CO_2}$ variation.** The central color panels are from Fig. 3. Side panels **A**-**D** show the climatic response of the planet, either inhabited ($T_{BioGeo}$, plain curves) or lifeless ($T_{Geo}$, dashed curves), to periodic variation in $p\mathrm{CO_2}$, depending on ecosystem composition (MG or AG+AT). The corresponding abiotic temperature variation amplitude is $\Delta T_{Geo}$ = 20 °C (indicated by the white dots and arrows in the central color panels). On a warm planet inhabited by MG ($T_{Geo}$ ranging from 30 to 50 °C), the climate response is buffered by about 20 % (A). However, on a cool planet ($T_{Geo}$ ranging from -20 to 0 °C, panel) the MG ecosystem amplifies the climate response by up to 33% (B). On a planet inhabited by AG+AT, the climate response to $p$CO2 variation is always amplified, but much less on a warm planet (C) (5% for $T_{Geo}$ ranging from 30 to 50 °C) than on a cool planet (33% for $T_{Geo}$ ranging from -10 to 10 °C, bottom-right panel) (D).

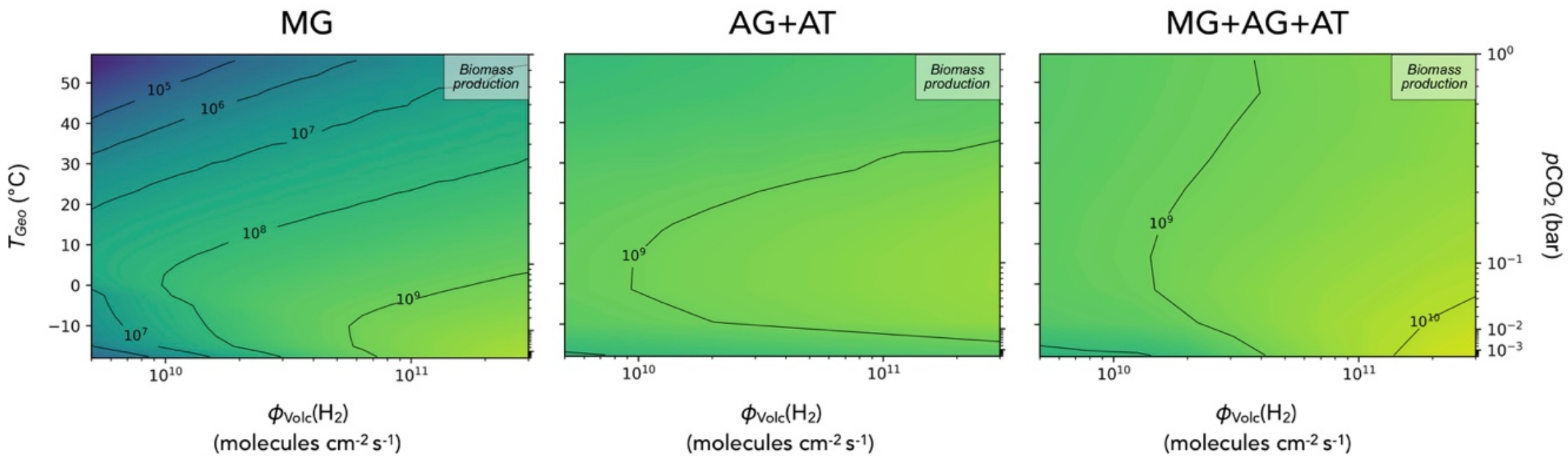

**Supplementary Figure 5.**
**Equilibrium biomass production (in molecules C cm$^{-2}$ s$^{-1}$) for each ecosystem composition.** The abiotic surface temperature, $T_{Geo}$, is determined by $pCO_2$. Other parameters are set to their default values (Supplementary Tables 2 and 3).

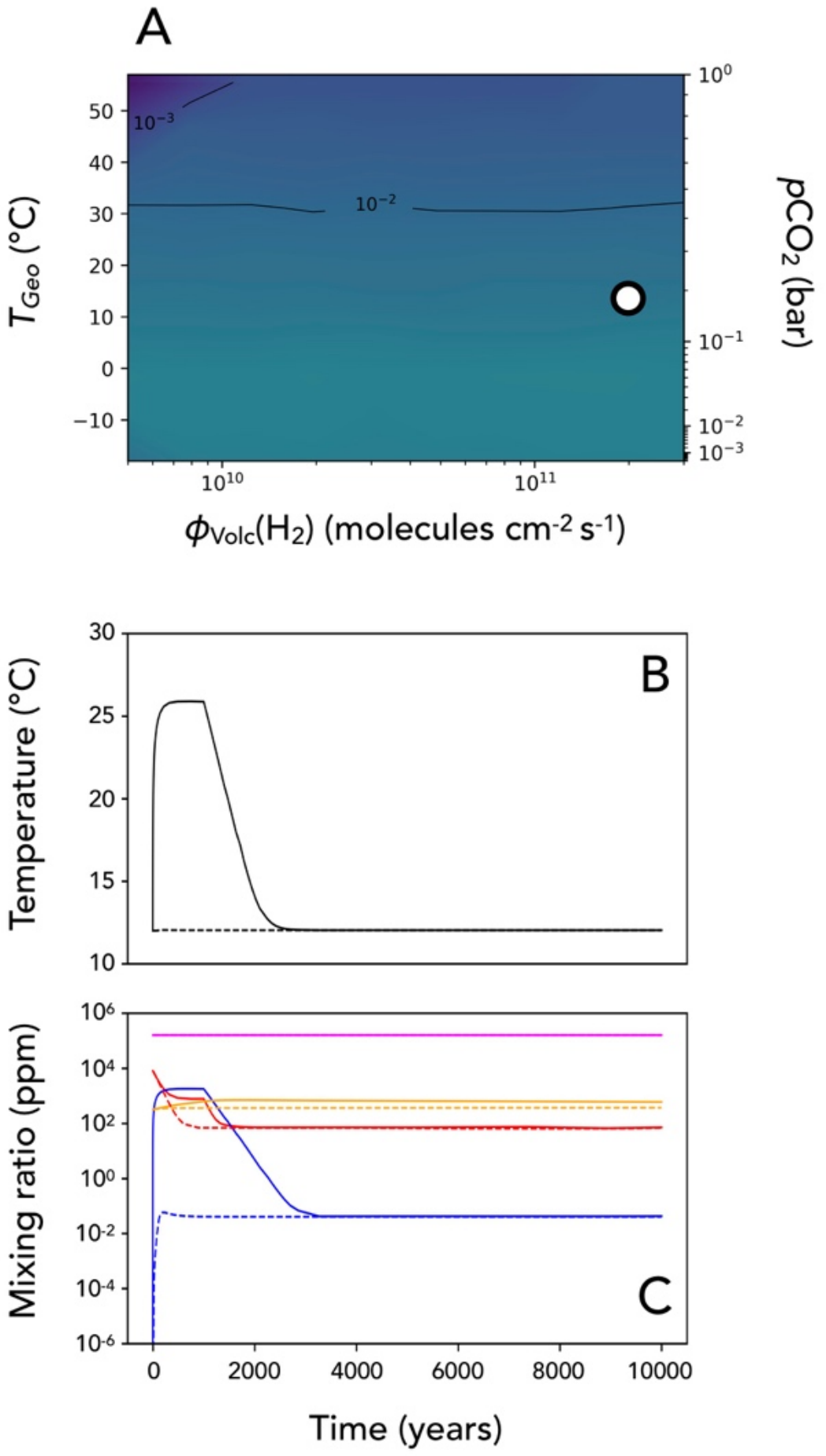

**Supplementary Figure 6.**

**Biogeochemical response of the early Archean Earth as MT evolves into the MG ecosystem.** (**A**) Effect of $H_2$ volcanic outgassing and abiotic temperature, $T_{Geo}$, on atmospheric $pCH_4$ at ecosystem-climate equilibrium. Here $T_{Geo}$ is determined by $pCO_2$. (**B**, **C**) Effect of MT evolving with MG (dotted lines) or 1,000 years after MG (plain lines) on temperature (B) and mixing ratios (C) of $CH_4$ (blue), $H_2$ (red), CO (yellow) and $CO_2$ (magenta), for $T_{Geo}$ and $\phi_{Volc}(H_2)$ indicated by the white dot in (A). All other parameters are fixed at their default values (Supplementary Tables 2 and 3).

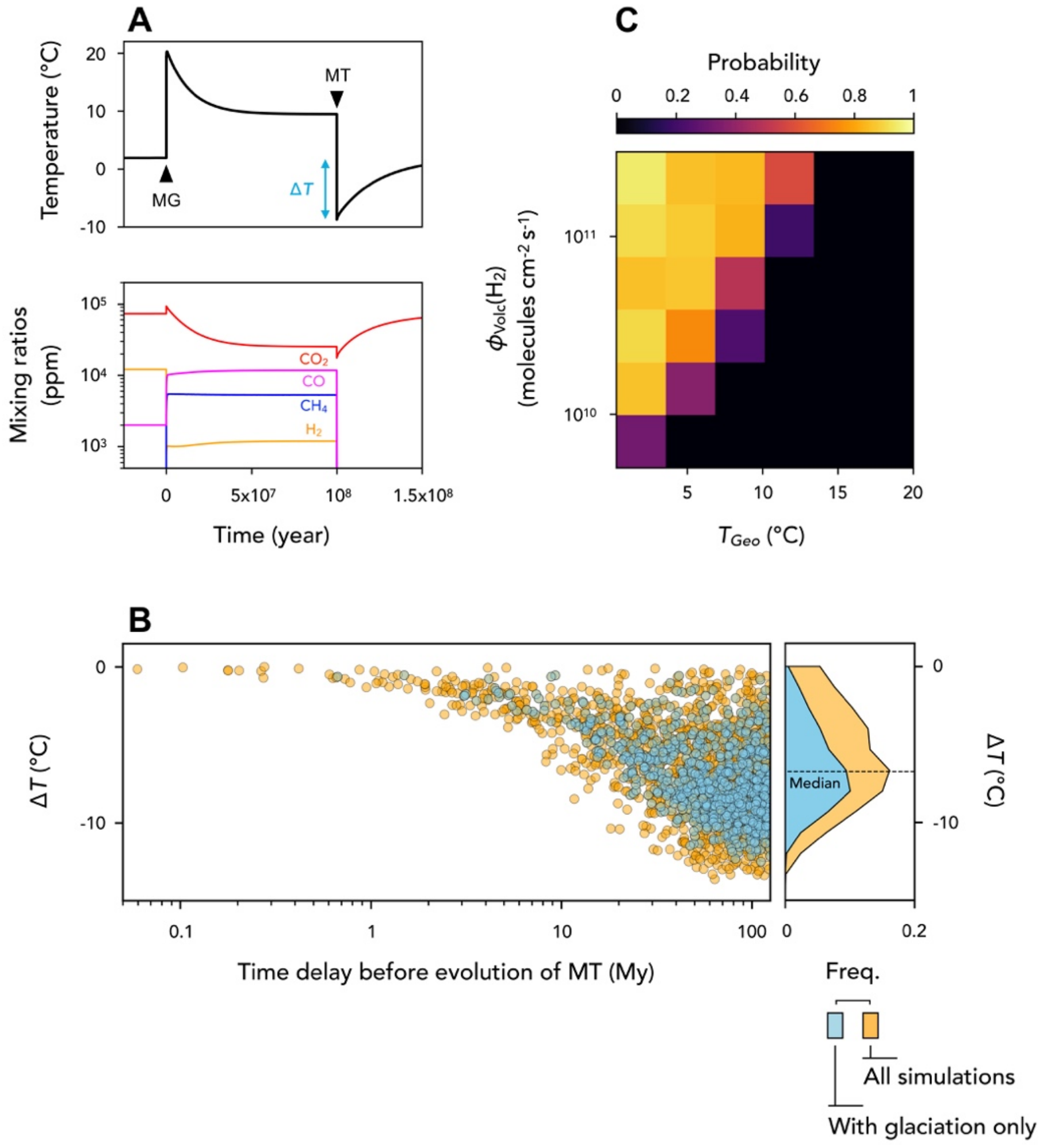

**Supplementary Figure 7.**

**Climatic and atmospheric destabilization by evolutionary metabolic innovation (methanotrophy).** (**A**) Example with $T_{Geo}$ = 2 °C, $\varphi_{volc}(H_2)$ = 3 $10^{11}$ molecules $s^{-1}$ $cm^{-2}$, and sulfur-based methanotrophs (MT) evolving 100 million years after MG (instead of MG-AG-AT as in the main text). *Top*, Change in surface temperature. *Bottom*, Change in atmospheric composition. Panels (**B**) Distribution of outcomes across a range of abiotic temperature $T_{Geo}$, $H_2$ volcanic flux, and evolution time of MT (2,000 randomly chosen combinations). *Left*, Amplitude of global cooling, $\Delta T$, with respect to the evolution time of MT. *Right*, Frequency distribution of all temperature changes $\Delta T$ (blue) and of temperature changes conditional on glaciation outcome (yellow). (**C**) Estimated probability of glaciation as a consequence of MT evolution, given the abiotic temperature $T_{Geo}$ and $H_2$ volcanic flux. Other parameters are set to their default values (Supplementary Tables 2 and 3).

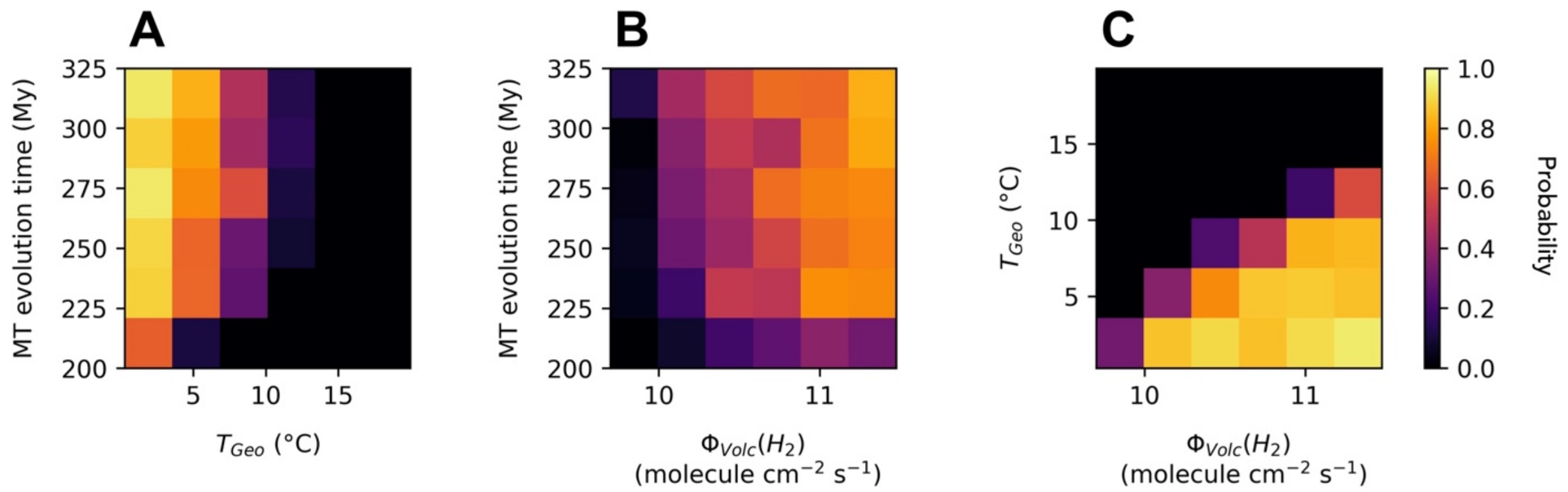

**Supplementary Figure 8.**
**Probability of climate destabilization (global cooling leading to glaciation) by the evolution of methanotrophy (MT).** Influence of three key parameters: (**A**) Abiotic temperature, $T_{Geo}$, and MT evolution time; (**B**) $H_2$ volcanic outgassing, $\phi_{Volc}(H_2)$**,** and MT evolution time; (**C**) $H_2$ volcanic outgassing, $\phi_{Volc}(H_2)$, and abiotic temperature, $T_{Geo}$.

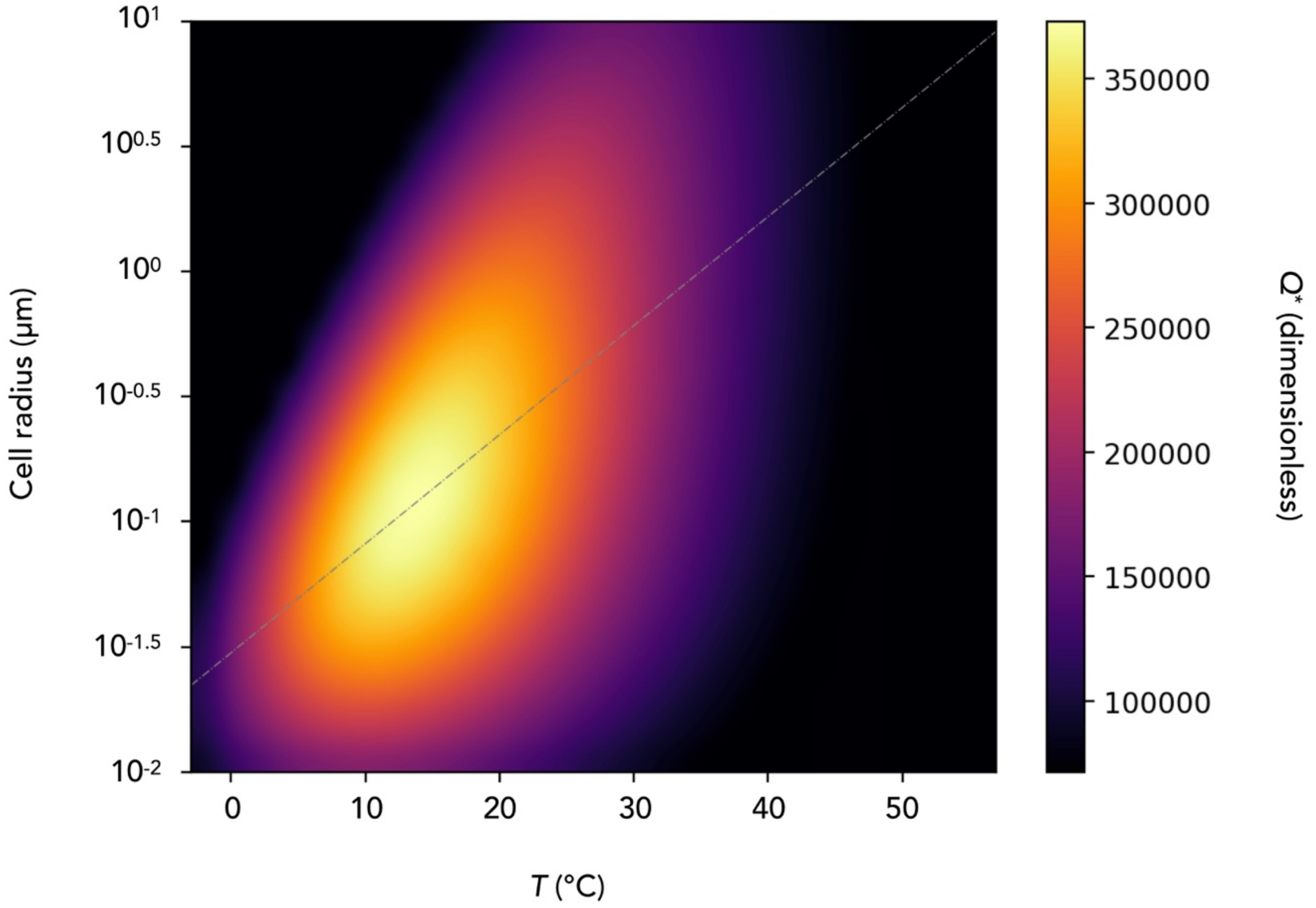

**Supplementary Figure 9.**
**Equilibrium thermodynamic reaction quotient, *Q**, (inversely correlated to resource use) as a function of cell radius and temperature.** The dashed line indicates the evolutionarily optimal cell size, i.e., cell size corresponding to the highest Q*, as a function of temperature. Parameters are set to their default values (Supplementary Tables 2 and 3).

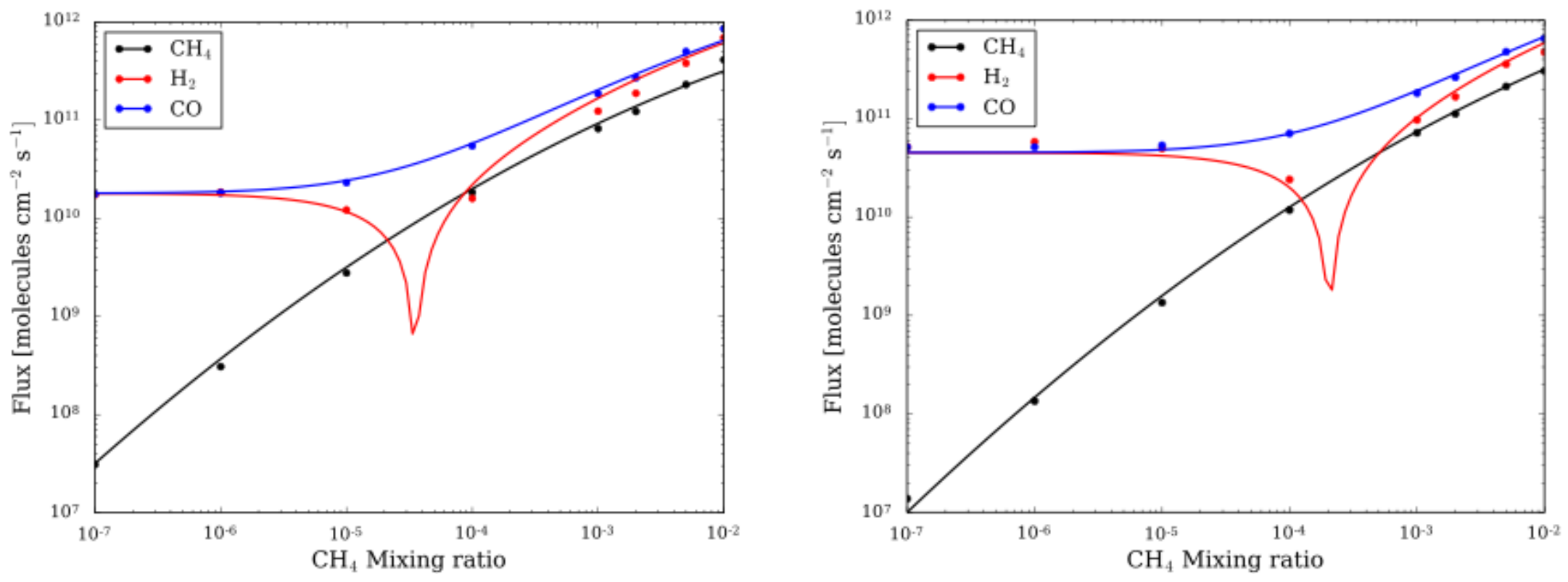

**Supplementary Figure 10.**

**Absolute net flux of $CH_4$, $H_2$ and $O_2$ due to $CO_2$ and $CH_4$ photolysis as a function of the $CH_4$ mixing ratio with 0.1 bar of $CO_2$, 100 ppm (*left*) and 1000 ppm of $H_2$ (*right*).** Dots correspond to the results of the 1D photochemical model and lines to our parameterization. The photochemistry leads to a production of CO and destruction of $CH_4$ in all cases. For low mixing ratios of $CH_4$, the $H_2$ flux is negative by the photolysis of $CO_2$. For high mixing ratios of $CH_4$, the $H_2$ flux is positive due to the photolysis of $CH_4$.

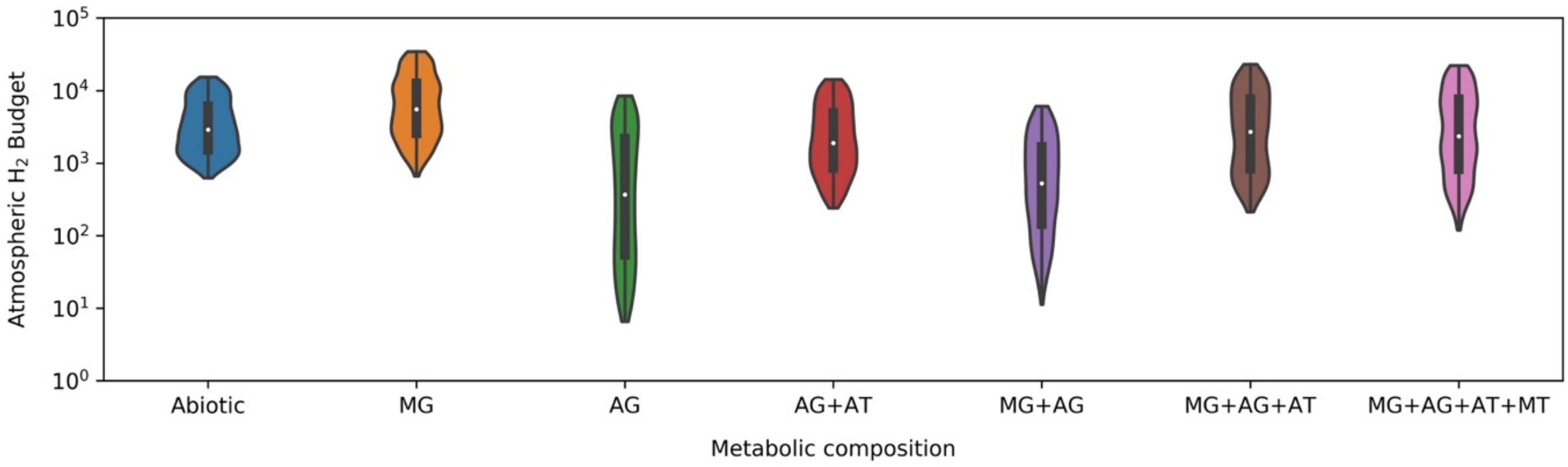

**Supplementary Figure 11.**

**Evolution of the atmospheric global redox budget evaluated as the atmospheric $H_2$ budget (in ppms) for each biosphere composition.** The global hydrogen budget of the atmosphere is computed as f($H_2$) + 4 f($CH_4$) + f(CO) following ref[6]. Results derived from the simulations presented in Fig. 4 (1,000 simulations in each scenario), obtained by drawing uniformly the model abiotic parameter values in log-uniform priors based on the litterature (see Table 1). The white dots represent the median of the distributions, the thick gray lines the interquartile range, and the thin gray lines the rest of the distribution.

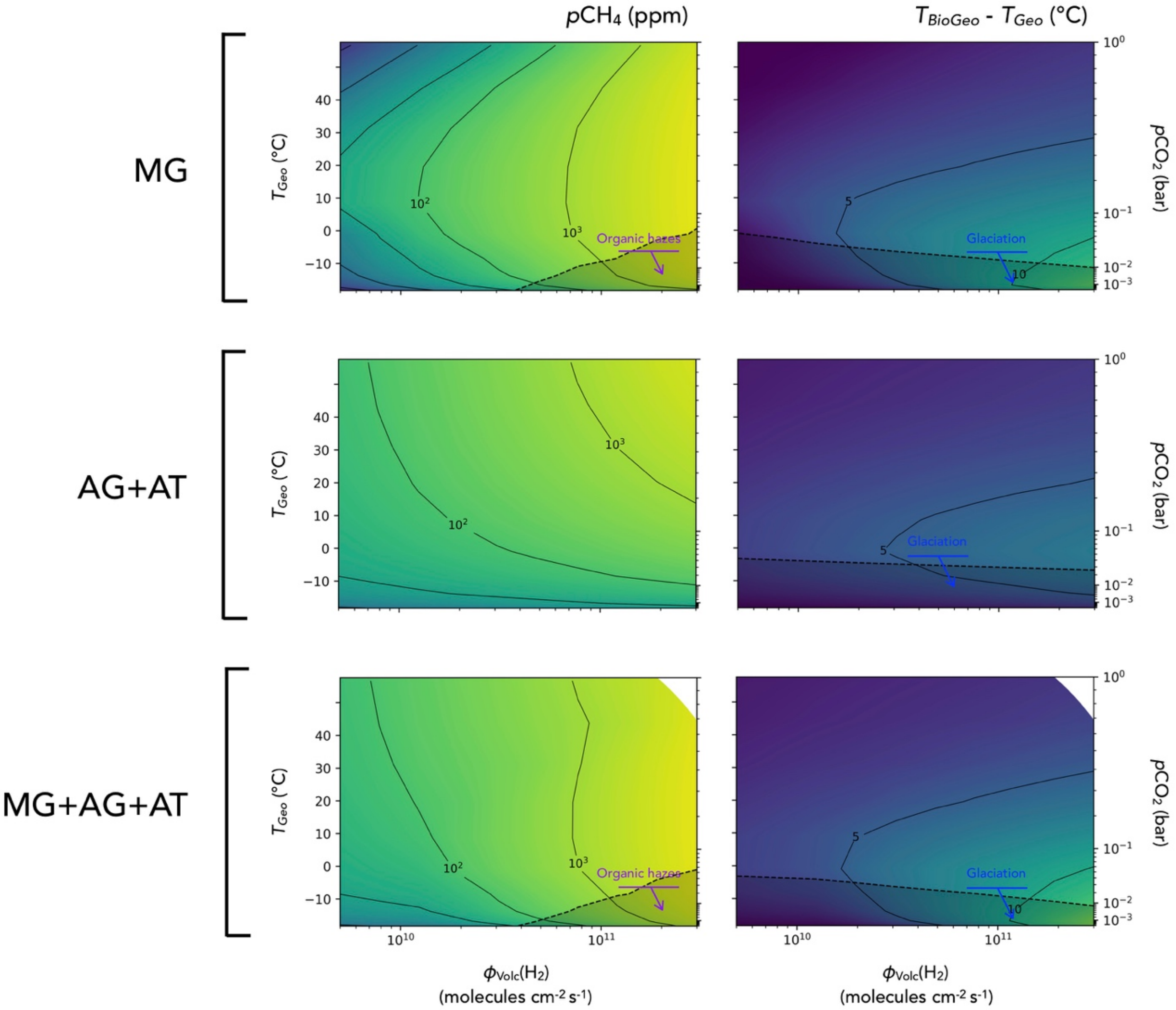

**Supplementary Figure 12.**

**Long-term effect of the carbon cycle on the biogeochemical response of the early Archean Earth to changes in $H_2$ volcanic outgassing, abiotic temperature ($T_{Geo}$) and ecosystem composition.** Here $T_{Geo}$ is determined by $pCO_2$ in the climate model. *Left*, Atmospheric $pCH_4$ at ecosystem-climate equilibrium. Shaded areas indicate conditions for organic haze formation ($pCH_4$:$pCO_2 > 0.2$). *Right*, Temperature differential between $T_{Geo}$ and the global surface temperature reached at ecosystem-climate equilibrium, $T_{BioGeo}$. Shaded areas indicate conditions leading to glaciation ($T_{BioGeo} < 0$°C). Other parameters are fixed to their default values (Supplementary Tables 2 and 3).

**Supplementary Table 1.**
**Metabolic reactions and their thermodynamic constants.**

| Name | Notation | Catabolic reaction | $\Delta G_0$ | $\Delta H_0$ |
|---|---|---|---|---|
| $H_2$-based methanogens | MG | $4 \cdot H_2 + CO_2 \rightarrow CH_4 + 2 \cdot H_2O$ | -32.6 $kJ.mol_{e^-}^{-1}$ | -63.2 $kJ.mol_{e^-}^{-1}$ |
| Sulfur-based methanotrophs | MT | $CH_4 + H_2SO_4 \rightarrow H_2S + CO_2 + 2 \cdot H_2O$ | -107 $kJ.mol_{e^-}^{-1}$ | -1.8 $kJ.mol_{e^-}^{-1}$ |
| Acetogens | AG | $4 \cdot CO + 2 \cdot H_2O \rightarrow CH_3COOH + 2 \cdot CO_2$ | -77.9 $kJ.mol_{e^-}^{-1}$ | -129.9 $kJ.mol_{e^-}^{-1}$ |
| Acetotrophs | AT | $CH_3COOH \rightarrow CH_4 + CO_2$ | -55 $kJ.mol_{e^-}^{-1}$ | 16.2 $kJ.mol_{e^-}^{-1}$ |
| Shared anabolic reaction: | | $10 \cdot CO_2 + N_2 + 24 \cdot H_2 \rightarrow C_{10}H_{18}O_5N_2 + 1.5 \cdot H_2O$ | 28.25 $kJ.mol_{e^-}^{-1}$ | 128 $kJ.mol_{e^-}^{-1}$ |

**Supplementary Table 2.**
**Default parameter values in the biological model.**

| Parameter (and reference) | Notation | Value or expression | Unit |
|---|---|---|---|
| Cell radius | $r_C$ | $10^{a_r+b_r \cdot T}$ | $\mu m$ |
| Cell volume | $V_C$ | $\frac{4}{3} \cdot \pi \cdot {r_c}^3$ | $\mu m^3$ |
| Structural carbon content | $B_{Struct}$ [17] | $18 \cdot 10^{-15} \cdot {V_C}^{0.94}$ | $mol\ C\ cell^{-1}$ |
| Maximum metabolic rate | $q_{max}$ | $e^{a_q+b_q \cdot T} \cdot {V_C}^{c_q}$ | $d^{-1}$ |
| | $a_q$ [18] | $-55.76$ | |
| | $b_q$ [19] | 0.1 | |
| | $c_q$ [21,22] | 0.82 | |
| Half-saturation constant | $K_S$ | $10^{-9}$ | $mol.L^{-1}$ |
| Maintenance rate | $E_m$ | $e^{a_E+b_E \cdot T} \cdot {V_C}^{c_E} \cdot 10^{-3}$ | $kJ.d^{-1}$ |
| | $a_E$ [18] | $-43.54$ | |
| | $b_E$ [20] | 0.08 | |
| | $c_E$ [23] | 0.67 | |
| Decay rate | $k_d$ | 0.5 | $d^{-1}$ |
| Basal mortality rate | $m$ | 0.1 | $d^{-1}$ |
| Maximum division rate | $r_{max}$ | 1 | $d^{-1}$ |
| Division rate dependence on internal reserve | $\theta$ | 10 | dimensionless |

**Supplementary Table 3.**
**Default values for parameters in metabolism specific size dependency on temperature.**

| Metabolism | $a_r$ | $b_r$ |
|---|---|---|
| MG | -13.23 | 0.0431 |
| MT | -13.289 | 0.0432 |
| AG | -13.21 | 0.044 |
| AT | -12.55 | 0.042 |

**Supplementary Table 4.**
**Ranges of biological parameter values used in Monte-Carlo simulations.**

| Parameter | Symbole | Range explored | Prior distribution |
|---|---|---|---|
| Maximum metabolic rate | $a_q$ | -55.20 – -56 | uniform |
| | $b_q$ | 0.076 – 0.12 | uniform |
| | $c_q$ | 0.53 – 1.10 | uniform |
| Half-saturation constant | $K_S$ | $10^{-11}$ – $10^{-6}$ | log-uniform |
| Maintenance rate | $a_E$ | -43.03 – -43.93 | uniform |
| | $b_E$ | 0.059 – 0.098 | uniform |
| | $c_E$ | 0.66 – 0.94 | uniform |
| Decay rate | $k_d$ | 0.05 – 5 | log-uniform |
| Basal mortality rate | $m$ | 0.01 – $r_{max}$ | log-uniform |
| Maximum division rate | $r_{max}$ | 0.1 – 10 | log-uniform |
| Division rate dependence on internal reserve | $\theta$ | 1 – 100 | log-uniform |

**Supplementary References**

1. Catling, David C., and James F. Kasting. *Atmospheric evolution on inhabited and lifeless worlds*. Cambridge University Press, 2017.

2. Kharecha, P., J. Kasting, and J. Siefert. "A coupled atmosphere–ecosystem model of the early Archean Earth." *Geobiology* 3.2 (2005): 53-76.

3. Kleerebezem, Robbert, and Mark CM Van Loosdrecht. "A generalized method for thermodynamic state analysis of environmental systems." *Critical Reviews in Environmental Science and Technology* 40.1 (2010): 1-54.

4. González-Cabaleiro, Rebeca, Juan M. Lema, and Jorge Rodríguez. "Metabolic energy-based modelling explains product yielding in anaerobic mixed culture fermentations." *PLoS One* 10.5 (2015): e0126739.

5. Tijhuis, L., Mark CM Van Loosdrecht, and J. J. Heijnen. "A thermodynamically based correlation for maintenance Gibbs energy requirements in aerobic and anaerobic chemotrophic growth." *Biotechnology and bioengineering* 42.4 (1993): 509-519.

6. Litchman, Elena, et al. "The role of functional traits and trade-offs in structuring phytoplankton communities: scaling from cellular to ecosystem level." *Ecology letters* 10.12 (2007): 1170-1181.

7. Ward, Ben A., et al. "The size dependence of phytoplankton growth rates: a trade-off between nutrient uptake and metabolism." *The American Naturalist* 189.2 (2017): 170-177.

8. Gillooly, James F., et al. "Effects of size and temperature on metabolic rate." *science* 293.5538 (2001): 2248-2251.

9. Aksnes, D. L., and J. K. Egge. "A theoretical model for nutrient uptake in phytoplankton." *Marine ecology progress series. Oldendorf* 70.1 (1991): 65-72.